\documentclass[11pt]{article}

\usepackage{amsmath,amsthm,amssymb,amscd,bbm,url}
\usepackage{authblk}
\usepackage{tikz}
\usepackage{subcaption}
\usepackage{booktabs}
\usepackage{xurl}
\usepackage{csquotes}

\usepackage[utf8]{inputenc}
\usepackage{geometry}
\usepackage{sectsty}
\usepackage[normalem]{ulem}

\sectionfont{\sffamily\bfseries\upshape\large}
\subsectionfont{\sffamily\bfseries\upshape\normalsize}
\subsubsectionfont{\sffamily\mdseries\upshape\normalsize}
\makeatletter
\renewcommand\@seccntformat[1]{\csname the#1\endcsname.\quad}

\makeatletter
\def\@maketitle{%
  \begin{center}%
  \let \footnote \thanks
    {\large \@title \par}%
    {\normalsize
      \begin{tabular}[t]{c}%
        \@author
      \end{tabular}\par}%
    {\small \@date}%
  \end{center}%
}
\makeatother

\newcommand{\R}{\mathbb{R}}
\newcommand{\prob}{\mathbb{P}}

\DeclareMathOperator*{\argmin}{arg\,min}

\makeatletter
\newcommand\newtag[2]{#1\def\@currentlabel{#1}\label{#2}}
\makeatother

\begin{document}
\title{\bf A calibrated BISG for inferring race from surname and geolocation}
\author{Philip Greengard\thanks{Department of Statistics, Columbia University, New York, pg2118@columbia.edu} \ and
Andrew Gelman\thanks{Department of Statistics and Department of Political Science, Columbia University, New York, ag389@columbia.edu}}

\date{9 June 2025}

\maketitle

\begin{abstract}
Bayesian Improved Surname Geocoding (BISG) is a ubiquitous tool for predicting race and ethnicity using an individual's geolocation and surname. Here we demonstrate that statistical dependence of surname and geolocation within racial/ethnic categories in the United States results in biases for minority subpopulations, and we introduce a raking-based improvement. Our method augments the data used by BISG---distributions of race by geolocation and race by surname---with the distribution of surname by geolocation obtained from state voter files. We validate our algorithm on state voter registration lists that contain self-identified race/ethnicity. 
\end{abstract}

\section{Introduction}
%
Accurately measuring racial bias or ``unfairness" in algorithms and systems is critical in many domains including machine learning, election law, data-driven predictions, lending practices, and healthcare coverage. However, a challenge to assessing and addressing racial disparities arises due to the absence of complete or reliable race and ethnicity information in many of these applications. To overcome this limitation, researchers and practitioners often use proxy methods that fill in, or impute, the missing data, allowing for racial disparities to be identified more effectively. 

Bayesian Improved Surname Geocoding (BISG) is a widely used algorithm for race/ethnicity 
imputation \cite{elliott2009} that gives a probabilistic estimate of race/ethnicity for an 
individual in the United States using surname and geolocation (often census tract or block group). 
BISG has been used to measure racial disparities across a range of applications 
in government, academia, and industry including health 
care \cite{elliott2009, haas2019}, lending practices \cite{cfpb2014}, 
voting patterns \cite{fraga2018}, user experience in social media 
\cite{meta_bisg}, and many other areas \cite{edwards2019, hepburn2020, studdert2020}. 
For example, the U.S. Department of Justice and the Consumer Finance Protection Bureau 
entered into a \$98 million settlement with Ally Financial for racially-biased 
lending, based in part on BISG \cite{cfpb2013}. Meta 
announced the launch of a BISG-based algorithm for reducing racial bias in 
advertisement delivery as part of a recent settlement 
with the Departments of Justice and Housing and Urban Development
\cite{meta_bisg2}. 
Recently, BISG was used to infer race/ethnicity of voters 
in federal voting rights cases, and the use of BISG was subsequently upheld in the 
U.S. Court of Appeals \cite{naacp_case, us_v_eastpointe, barreto2022, decter-frain2022}.
\begin{figure}
\centering
\begin{subfigure}[b]{0.99\linewidth}
\centering
  \includegraphics[height=8.2cm, keepaspectratio]{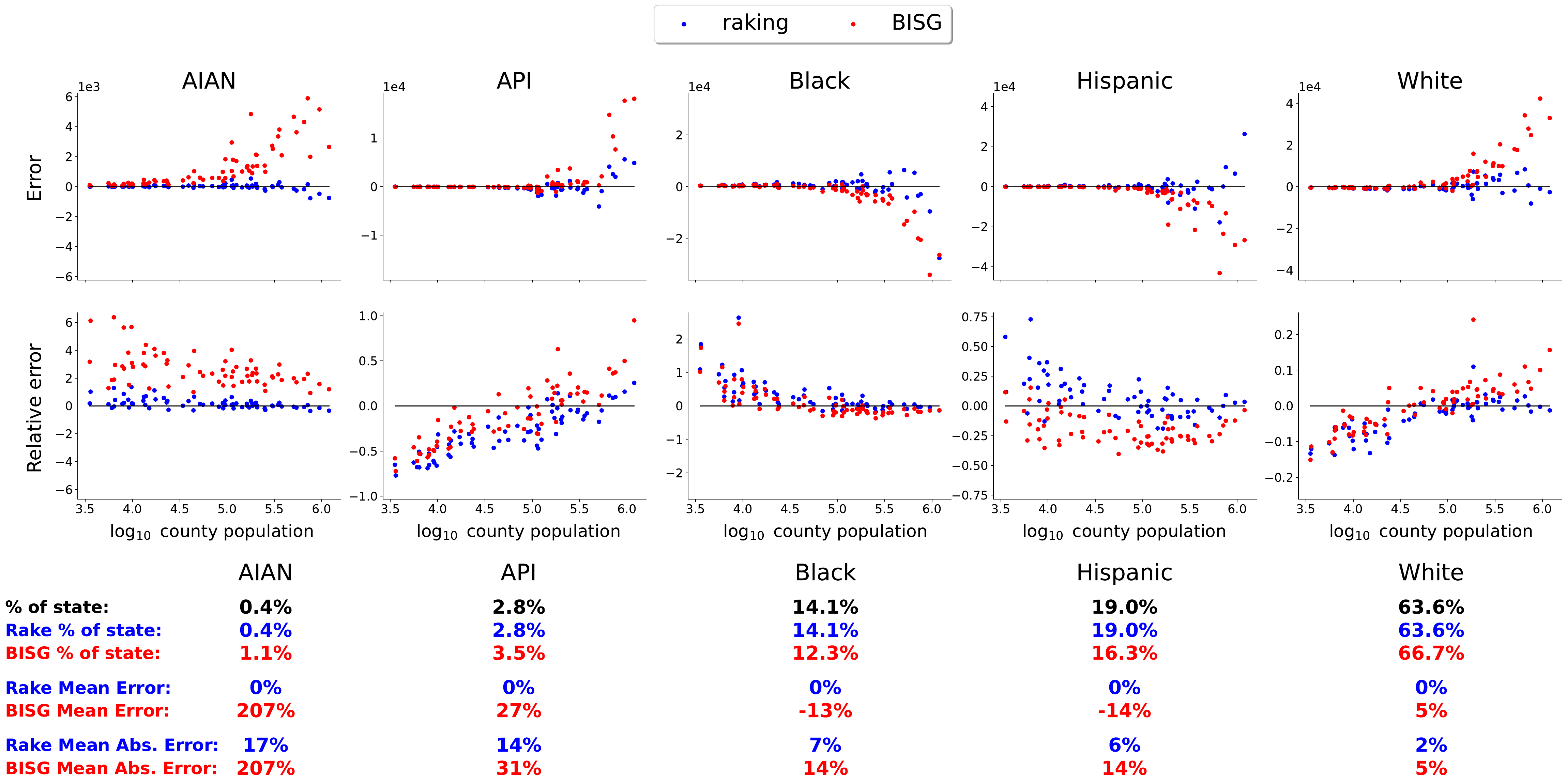}
\end{subfigure}
  \caption{\em Errors in subpopulation estimation for subsampled registered voters in each county in Florida in 2020 for two methods, BISG and raking. Each dot in the scatterplots represents error in one county with BISG or raking. Dots above the horizontal line correspond to overestimating the size of a subpopulation size in one county. Dots below the horizontal line indicate underestimating.}
    \label{fig:p1}
\end{figure}

BISG has become popular because it uses publicly available census 
data and is mathematically straightforward. However, BISG relies on a crucial 
assumption---that, within the U.S. population, 
surname and geolocation are conditionally independent of race. This means that 
if we know a person's race, then their geolocation tells us nothing 
about their surname, and likewise, their surname gives no 
information about their geolocation.
While this assumption may be mathematically and practically convenient, it appears 
to contradict the conventional wisdom that people tend to live in close 
proximity to their relatives and others with similar demographic features.\footnote{These and other likely 
sources of inaccuracy in BISG's conditional independence assumption are noted in \cite{imai2016, imai2022}.}
For example, 
Korean surnames are prevalent in Fort Lee, New Jersey, while 
across the river in Manhattan, Asian surnames are much more likely to be Chinese.

Here, we demonstrate that dependence of surname and geolocation within 
racial/ethnic groups results in systematic inaccuracies in BISG, and we introduce a method for improving the predictions. 
We analyze the impact of BISG's independence assumption using 
voter registration lists in Florida and North Carolina, two of a handful of 
U.S. states that provide labeled race/ethnicity information on registered voters. 
Using state voter files, we construct a fully self-consistent BISG with exact
factors and validate its accuracy on the full population on which it is trained.  
We demonstrate empirically that BISG consistently suffers from poor 
relative accuracy for small subpopulations. 
Furthermore, within subpopulations, the sign of the errors is consistent. 
For example, in Florida 
BISG underestimated the size of Hispanic subpopulations in $88\%$ of counties 
and in nearly $50\%$ of counties, Hispanic subpopulations were underestimated 
by at least $25\%$ (see Figure \ref{fig:fl_vf_scatter}). Similarly, the size of 
Asian and Pacific Islander (API) subpopulations was underestimated in $90\%$ of counties
and in $40\%$ of counties, API subpopulations were underestimated by at least $25\%$. 
We provide extensive results of this study in Appendix \ref{a:bisg_fl}. 

The prediction method we introduce is straightforward to implement, 
easily interpretable, and is more accurate and better calibrated than BISG.
Our algorithm uses the inputs to BISG (joint distributions of race by 
geolocation and race by surname) in combination 
with two new data sources---the joint distribution of surname and geolocation obtained from state voter 
registration lists and the statewide race/ethnicity distribution of registered voters 
from the Voter Supplement of the Current Population 
Survey (CPS). Our predictions can be computed by first evaluating BISG 
predictions for all registered voters in a state and then modifying those
predictions so that they coincide with known margins. This procedure is 
done via raking~\cite{deming1940}, a classical and widely used algorithm for fitting 
observed data to known margins.
Summing our predictions over all surnames and geolocations in a given 
state gives a race/ethnicity distribution that matches the distribution estimated by the CPS. 
Similarly, our predictions' surname-by-geolocation two-way margin (summing over
race/ethnicity) matches the margins of state voter registration lists. 
The accuracy of our raking predictions depends on the accuracy and consistency of the 
margins that we rake to. These margins that have been widely used and studied in political science
literature including their consistency and accuracy \cite{ghitza2020, mcdonald2007, ansolabehere2022, fabina2022}. 

If one were to sum BISG estimates over everyone in a particular state, 
the race/ethnicity estimate for that state would not match the true distribution. 
This may come as surprise, given that this information can be obtained from the 
race by geolocation inputs into BISG. 
However, without the joint distribution of surname and geolocation, BISG has no 
mechanism to enforce the accuracy of the statewide race/ethnicity margin. 
In the states we have tested, BISG's statewide race/ethnicity margins are often 
inaccurate, and county-level errors are frequently larger (see, e.g., Figure \ref{fig:p1}). 
A main advantage of our approach is that we use the joint distribution of 
surname and geolocation in state voter files to ensure that  
statewide margins of our estimates are accurate, at least up to the accuracy 
of state voter files and the Current Population Survey, two widely used and studied data sets. 

Much of the literature on BISG focuses on error in measuring racial disparity on 
some outcome of interest. 
That is, the goal is to measure the race/ethnicity distribution by, for instance, 
political party, for a group of people with unknown race/ethnicity. 
Often, BISG is used to impute race/ethnicity of those people, and standard methods, 
or recently introduced alternatives such as \cite{kallus2022, mccartan2023}, are 
subsequently used to estimate racial disparities.\footnote{In many applications, 
even when mean errors in BISG 
are negligible over the population of interest, standard methods for disparity 
analysis are biased. In particular, when BISG errors are correlated 
with an outcome of interest, bias is introduced in disparity estimates~\cite{chen2019}. 
Techniques such as \cite{mccartan2023, kallus2022} address these biases.} 
The accuracy of these racial disparity estimates combines various sources of error, 
including BISG error, correlation of BISG error with the outcome of interest, and 
the assumptions of the method used for disparity analysis~\cite{mccartan2023}. 
There is a vast range of downstream applications of BISG to racial 
disparity measurement, and notions of errors in those applications 
\cite{airbnb, cfpb2014, meta_bisg, elliott2009, edwards2019, hepburn2020, studdert2020}.
We focus on classes of errors that are likely to impact many applications, albeit
possibly in different ways for different applications. 
Our method and BISG are algorithms for estimating the joint distribution of race/ethnicity, 
surname, and geolocation and we use a variety of metrics to compare BISG and our 
method to the true joint distribution of race/ethnicity, surname, and geolocation. 
The subpopulation errors we report---estimated race/ethnicity of registered voters in various geographic
regions (see, e.g., 
Figures \ref{fig:p1}, \ref{fig:nc2020_scatter_bar_sub}, and \ref{fig:nc2010_scatter_bar_sub})---are also of applied interest in electoral politics, election law, 
and political science
\cite{deluca2023, barreto2022, naacp_case, us_v_eastpointe, henderson2016, 
barreto2004}. 

We estimate race/ethnicity distributions of registered voters in Florida and North 
Carolina and report individual-level and mean errors across geographic regions. 
We use two methods 
to calibrate our test set (registered voters with labeled race/ethnicity) to the CPS 
margin: 
(i) subsampling the test set to match the CPS race/ethnicity distribution and 
(ii) transforming predictions from the voter file population 
to the CPS-implied population. 

Our raking algorithm outperforms BISG across a range of metrics 
and both calibration approaches. 
For example, for the subsampled voter file, the mean absolute error 
of raking predictions in Florida in 
2020 is $4\%$ compared to $10\%$ for BISG. In North Carolina, raking 
mean absolute error is $3\%$ compared to $7\%$ for BISG in 2020
and $3\%$ for raking versus $5\%$ for BISG in 2010. 
In Figure \ref{fig:p1}, we plot the accuracy of estimates of subpopulation size 
implied by BISG and raking in all counties in Florida in 2020. 
Figures \ref{fig:nc2020_scatter_bar_sub} and \ref{fig:nc2010_scatter_bar_sub} 
include the same results for North Carolina in 2020 and 2010 respectively. 

Figures \ref{fig:p1}, \ref{fig:nc2020_scatter_bar_sub}, and \ref{fig:nc2010_scatter_bar_sub}
illustrate that mean errors of BISG are often large and correlated with county population.\footnote{Similarly, errors are also correlated with subpopulation size. For example, Hispanics are consistently undercounted in counties with small Hispanic populations in Florida in 2020. There are many features strongly correlated with county population in Florida and North Carolina and thus many features correlated with BISG errors in those states. We do not investigate those correlations in this paper. Recent work has shown that BISG errors are correlated with various socioeconomic and demographic factors \cite{argyle2023}.} For example, in Florida 
in 2020, BISG overestimates the size of API subpopulations by over $25\%$ while 
American Indian and Alaskan Native (AIAN) subpopulations are overestimated by over 
$100\%$. Our raking predictions, by design, recover the exact size of each 
subpopulation over the state. However, the errors of our method are also 
correlated with county population size. 

In addition to validating our method and BISG in Florida and North Carolina, 
we implemented our prediction algorithm and BISG in several states 
without publicly available race/ethnicity information---New York, Ohio, Oklahoma, 
Vermont, and Washington. Our predictions and implementation for these states as 
well as Florida and North Carolina are publicly available.\footnote{ 
\url{https://github.com/pgree/raking_bisg}} 

BISG was originally introduced for healthcare applications~\cite{elliott2009},
and public health researchers have subsequently made methodological 
progress on race/ethnicity imputation. For example, Medicare BISG \cite{martino2013, 
haas2019} relaxes the independence assumption and uses logistic regression on 
demographic features to predict race/ethnicity. 
The public health family of methods and the applications they address are, 
from a statistical standpoint, distinct from research on BISG for registered 
voters or the general population. The primary difference is that in healthcare 
applications, practitioners often have access to large amounts of labeled data---data sets that 
include surname, geolocation, race/ethnicity, and other demographic data 
for a large percentage of the population of interest~\cite{elliott2009, haas2019}. 
Unfortunately there is no equivalent for the registered voter population outside 
of a few state voter registration lists. It is possible that methods like Medicare 
BISG can be applied to those voter registration lists, though we do not investigate 
that in this paper. 

The central topics of this paper---a simple improvement on 
BISG and an analysis of BISG's errors---have been the subject of 
a wide literature. For example, failure modes of BISG are discussed in 
\cite{chen2019, kallus2022, baines2014, zhang2018, haas2022, argyle2023} and 
race/ethnicity prediction algorithms include BISG-based 
methods \cite{imai2016, imai2022, haas2019} and many others 
\cite{jain2022, sood2018, lee2017}. 
This paper is distinct from those in two primary ways: 
(i) we introduce a simple and easily interpretable algorithm, which by
itself is a major improvement on BISG 
and can also be combined with other race/ethnicity prediction algorithms, 
including several of the aforementioned ones,
and (ii) we provide a novel approach to error analysis---isolating the impact of
BISG's independence assumption by computing
BISG predictions on a complete and self-contained data set with exact factors.
We include a longer discussion of the relevant literature in Appendix \ref{a:lit_review}. 

The remainder of this paper is organized as follows. Section \ref{s:method}
describes the raking-based prediction method that we propose. In Section \ref{s:results},
we present the results of validation of our method on predicting the race/ethnicity of 
registered voters in Florida and  North Carolina. We discuss the results of our validation
and directions for future work in Section \ref{s:conclusion}. In the appendices, we 
first provide detailed results of our analysis of BISG on self-contained and complete 
data sets of registered voters (Appendix \ref{a:bisg_fl}). Appendix \ref{a:bisg_voters}
details the BISG prediction that we use for the results in Section \ref{s:results}. 
In Appendix \ref{a:calib_map}, we describe an approach to validation of voter registration
lists that uses a calibration map. Appendix \ref{a:race_cat} contains details on combining 
race/ethnicity classifications from the various data sets used in this analysis.
Detailed results from validation on the accuracy and calibration of BISG and raking is 
contained in Appendix \ref{a:details}, and Appendix \ref{a:lit_review} provides a literature review.

\section{Methods}\label{s:method}
In this section, we propose an algorithm for predicting race/ethnicity that uses 
raking (see, e.g., \cite{deming1940}) in combination with survey data
and state voter files to improve BISG predictions. 

We use the conventions and notation of log-linear modeling of \cite{bishop1975} 
for formulating our proposal. In this paper we focus on prediction of entries of 
three-way contingency tables of registered voters with dimensions surname, 
geolocation, and race/ethnicity. We denote by $x_{sgr}$ the known, correct entries 
of the three-way table of registered voters in a given state. 
The index $sgr$ corresponds to the $s^{\text{th}}$ surname, $g^{\text{th}}$
geolocation, and $r^{\text{th}}$ race/ethnicity
for $s = 1,\dots,n_s$, $g=1,\dots,n_g$, 
and $r=1,\dots,n_r$, where $n_s$ denotes the number of surnames
in the census list, $n_g$ the number of geographic regions (e.g., county), 
and $n_r$ the number of races/ethnicities (usually six).
We denote by $m_{sgr}$ predictions of $x_{sgr}$. In accordance with
\cite{bishop1975} we denote summing over components of $x_{sgr}$ with 
a $+$ subscript. So $x_{+gr} = \sum_{s} x_{sgr}$ is the two-way table of 
race by geolocation and 
$x_{+g+} = \sum_{s} \sum_{r} x_{sgr}$ is the vector of geolocations. 

In the usual formulation of BISG, predictions are normalized such that for any 
$(s, g)$ pair the sum of predictions over race is one. A trivial modification of the 
BISG formula instead results in estimates of the number of people in each cell 
$x_{sgr}$.
For convenience and readability, we use this formula of BISG, which can be 
expressed as 
\begin{align}\label{bisg_tau}
m_{sgr} = \frac{x^*_{+gr} x^*_{s+r}}{x^*_{++r}},
\end{align} 
where $x^*_{sgr}$ denotes the joint distribution of surname, 
geolocation, and race/ethnicity of the U.S. population. 
The full joint distribution $x^*_{sgr}$ of the U.S. population is not 
publicly available, but the margins that appear in equation \eqref{bisg_tau} are provided 
by the Census. 
%
%
We propose estimating $x_{sgr}$ with a prediction of the form 
\begin{align}\label{rake_model1}
m_{sgr} = x^*_{+gr} \, x^*_{s+r} \, x_{sg+} \exp(\theta_{r} + \theta_{sg})
\end{align}
where $x_{sg+}$ denotes the joint distribution of surname and geolocation 
of registered voters. 
The terms $\theta_{r}, \theta_{sg}$ are fit via raking such that our predictions' 
race/ethnicity margin $m_{++r}$ and surname by 
geolocation margin $m_{+sg}$ match known totals. 
A key difference between the parametric form of our proposal, \eqref{rake_model1},
and BISG, \eqref{bisg_tau}, is the term $x_{sg+}$, the joint distribution of surname 
and geolocation. The presence of this term means that our raking estimates, unlike BISG,
do not assume that geolocation and surname are independent conditional on 
race. The term $x_{sg+}$ also allows us to evaluate, and correct via raking, the 
prediction's marginal distribution of race/ethnicity in a state.

Raking predictions for registered voters can be computed by first evaluating 
BISG predictions for the all registered voters in the state, and then raking those 
predictions to the known margins $x_{++r}$ and $x_{sg+}$. 
The resulting estimates, $m_{sgr}$ in \eqref{rake_model1}, have the property that their
margins coincide with the known margins. That is, 
\begin{align}
m_{++r} = x_{++r} \quad \text{ and } \quad m_{sg+} = x_{sg+}.
\end{align}
For registered voters, the margin $x_{sg+}$ can be obtained from publicly
available voter registration lists and the racial breakdown of the 
registered voters in a state can be obtained from the
Voter Supplement of the Current Population Survey. 

The parametric form of our predictions, \eqref{rake_model1}, was chosen 
based on availability of accurate margins $x_{++r}, x_{sg+}$. Unfortunately for 
the accuracy of our predictions,
the race by geolocation margin $x_{+gr}$ is generally unavailable for registered 
voters. If it were available, we would also rake to that margin and 
\eqref{rake_model1} would include the factor $\exp(\theta_{rg})$. 
In other applications, however, another set of margins may be known, and thus
a parametric form other than \eqref{rake_model1} could be used. 

The predictions obtained through raking, such as in \eqref{rake_model1}, have 
theoretical properties that have been well-studied \cite{bishop1975}. 
For example, \eqref{rake_model1} are the unique predictions of the specified parametric
form that coincide with the known margins. They also minimize the KL-divergence
to the BISG estimates among all predictions that coincide with the known 
margins~\cite{darroch1972}.

\section{Results}\label{s:results}
We validate the proposed method on
predicting race/ethnicity of registered voters in Florida and North Carolina. Both 
of these states provide a publicly available data set that includes name, address, and 
self-identified race/ethnicity of registered voters. 
The registered voter lists include relevant information on voters that we do not use 
in this paper. That information is used in prediction methods such as \cite{imai2016, haas2019} 
and can be combined with the methods of this paper.

We make predictions using BISG and raking for the three-way contingency table $x_{sgr}$ for 
registered voters in Florida and North Carolina with dimensions surname, geolocation, 
and race. For geolocation, we use a county-level discretization.\footnote{Recent work has shown that the accuracy of BISG can benefit from 
finer discretizations than county \cite{clark2022}. The raking approach of this paper
can be used with any geographic discretization, though we leave validation of 
discretizations other than county to future work.}
For race/ethnicity, we use categories American Indian and Alaskan native (AIAN),
Asian and Pacific Islander (API), non-Hispanic Black, Hispanic, non-Hispanic White, 
and other. 

Predictions using both BISG and our proposal assume a distribution on race/ethnicity
of registered voters that is estimated in the Voter Supplement of the Current 
Population Survey (CPS), which is collected in November of congressional election years. 
The race/ethnicity distribution estimated by survey data differs from the labeled 
race/ethnicity of state-provided voter file lists. In some cases, such as Florida in 2020, 
the differences are minor, whereas in others, such as North Carolina in 2010 and 2020, 
differences are larger. 
Differences between the CPS survey data and voter file lists are due to several factors: 
sampling error, nonresponse in both survey and voter lists, designations of race/ethnicity 
(see Appendix \ref{a:race_cat}), and slightly different timeframes for collection of data.
For performing validation on voter registration lists in North Carolina and Florida, we use
two strategies for calibrating the test set (registered voters) with the population on which
the predictions were trained (CPS). In one approach, we
subsample voter file lists so that the race/ethnicity distribution of the test set (registered
voters) coincides with the CPS estimate of the race/ethnicity distribution of registered
voters. In the other approach, we make predictions on the full set of 
registered voters in the state (using the CPS-assumed distribution of race/ethnicity) 
and then map those predictions onto the race/ethnicity distribution of the voter file
using a calibration map. Results of validation using the calibration map are presented
in Appendix \ref{a:details} and the construction of the calibration map is described in 
Appendix \ref{a:calib_map}.

\begin{figure}
\centering
\begin{subfigure}[b]{0.99\linewidth}
\centering
  \includegraphics[height=8.2cm, keepaspectratio]{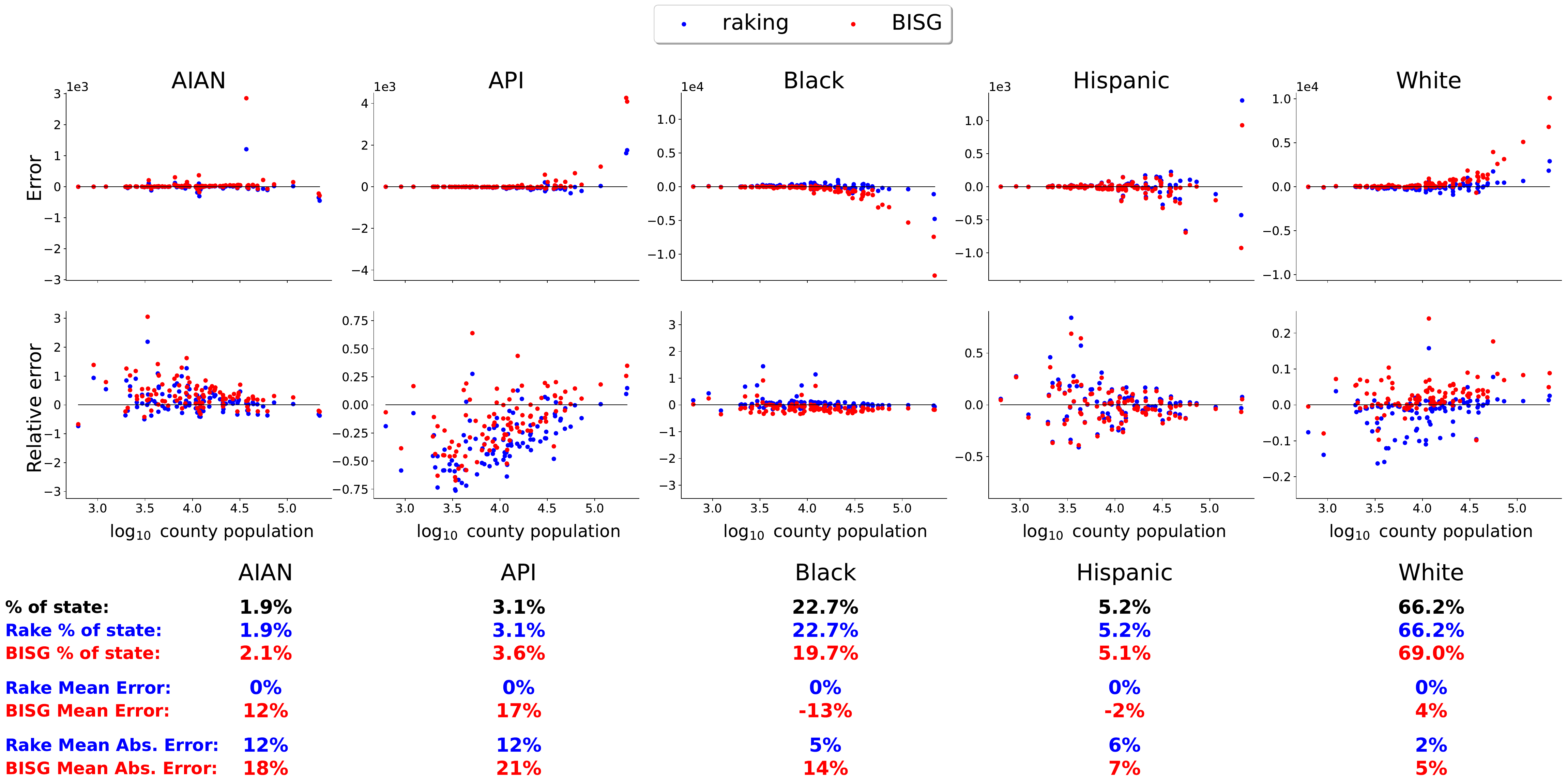}
\end{subfigure}
  \caption{\em Errors in subpopulation estimation for subsampled registered voters in each county in North Carolina in 2020 for two methods, BISG and raking. Each dot in the scatterplots represents error in one county. Dots above the horizontal line correspond to overestimating the size of a subpopulation size in one county. Dots below the horizontal line indicate underestimating.}
    \label{fig:nc2020_scatter_bar_sub}
\end{figure}

We compare the accuracy of our prediction algorithm, described in Section \ref{s:method}, 
to BISG. While BISG is often implemented by obtaining its factors (probability of 
geolocation given race and probability of race given surname)
from census data from the full population, we adjust the usual
BISG factors to correspond to the population of registered voters in Florida and North 
Carolina using the 
strategy of \cite{imai2016}. 
We provide details on our BISG implementation in Appendix \ref{a:bisg_voters}.  
We evaluate performance using two categories of metrics, subpopulation estimation
(or mean errors) and individual-level errors. 
To evaluate subpopulation estimates in a county, we first compute race/ethnicity
predictions for each registered voter in that county, and then sum the 
estimates (vectors in $\R^6$) over each person. This procedure, 
sometimes known as weighting or a weighted estimator~\cite{chen2019}, 
results in an estimate, under a prediction scheme, of the race/ethnicity 
distribution of the registered voters in the county. We then compare that distribution 
to the true distribution of the county, obtained from the labeled data set. We report
these errors in Figures \ref{fig:p1}, \ref{fig:nc2020_scatter_bar_sub}, and \ref{fig:nc2010_scatter_bar_sub}. 
Details on the error metrics are provided in Appendix \ref{a:details}. 

Individual-level errors compare a prediction for particular surname, geolocation pairs 
(e.g., Martinez, Miami-Dade County) to the empirical, true distribution. 
We compare the prediction
and ground truth (vectors in $\R^6$) in $\ell^1$, $\ell^2$, and we evaluate the 
negative log-likelihood under a multinomial model. We report these error metrics
aggregated at the region level in Table \ref{t:vf_region}.

While predictions in Florida and North Carolina are for validation purposes (we have
the self-identified race of registered voters), in other states, we have the 
surname-by-geolocation
and race/ethnicity margins, but no race/ethnicity labels for each registered voter. 
We implement our algorithm and BISG (in addition to several other predictions) in 
several such states: New York, Ohio, Oklahoma, Vermont, and Washington.\footnote{ 
An implementation and predictions are available at \url{https://github.com/pgree/raking_bisg/}.}

\section{Discussion and conclusion}\label{s:conclusion}
In this paper, we have shown that statistical dependence of surname and location
by race/ethnicity results in systematic inaccuracies in BISG. 
We have proposed a novel algorithm for predicting the joint distribution of race/ethnicity, geolocation, and surname. 
Our algorithm relaxes BISG's independence assumption, is mathematically
straightforward, easily interpretable, and noticeably more accurate than BISG. 
Instead of viewing BISG as a race/ethnicity imputation algorithm for individual 
surname and geolocation pairs, we view it as estimating entries of a 
contingency table with dimensions race/ethnicity, geolocation, and surname. 
We use raking, a classical tool in the analysis of tabular data \cite{bishop1975}, 
to combine the inputs to BISG with publicly available margins. 

By using the joint distribution of surname and geolocation from state voter files
in combination with survey data, our method produces an accurate estimate of 
the statewide distribution of race/ethnicity. 
This is a property that BISG does not have at any geographic region. 
With BISG, statewide mean errors are generally large and errors are often correlated 
with total county population (and subpopulation size), which can introduce bias in racial disparity estimation~\cite{chen2019}. 
We validate our method on predicting race/ethnicity of registered 
voters in Florida and North Carolina. Unlike BISG, the proposed method benefits from 
being unbiased in the sense that for any race/ethnicity, the mean error 
across all registered voters in a state is zero. 
Our method is also better calibrated than BISG. The Kuiper metrics for 
measuring calibration (Table \ref{t:kuiper}) demonstrate clear 
improvement \cite{tygert2021}.\footnote{The advantages of the graphical 
representation of calibration in Figure \ref{fig:calib_grid} and the Kuiper
metric are discussed in depth in \cite{arrieta2022}.}
On the other hand, like with BISG, the errors of our method are usually correlated
with county population size. 

In this paper we validate and fit our method on the same population---registered voters. 
However, our algorithm, like BISG, can be applied to test sets that differ from the data 
set it was fit on. BISG is used in a wide range of applications, including
for enrollees in health plans \cite{elliott2009}, users of social media \cite{meta_bisg}, 
and registered voters \cite{naacp_case}. In all of these cases, BISG is applied to 
populations different from the one it was fit on---the full U.S. population. 
It is possible to account for differences between the train and test populations
using adjustments like the logistic regression approach of \cite{elliott2009} or 
the approach of \cite{imai2016}, which we use in this paper
(see Appendix \ref{a:bisg_voters} for details). 
Similar approaches could be used to adjust this paper's raking-based estimates 
trained on the registered voter population. 
Comparing our method and BISG on application domains other than registered 
voters is an area of future study. 

There are many natural generalizations of this work. The methods of 
\cite{haas2019} include BISG generalizations for healthcare 
applications that use logistic regression for incorporating 
additional features. Similar strategies could be fruitful for improving accuracy
of our method. These methods would likely involve training on labeled data in 
certain geographic regions to generalize to unlabeled data in other areas.
Other potential generalizations include raking on 
the logistic scale, and the use of additional features for training, 
such as sex, first name, etc.

\section*{Acknowledgements}
We thank Yair Ghitza for suggesting the use of Florida's voter registration list to test BISG and for feedback on the project; Mark Tygert for pointing out potential inaccuracies in BISG's independence assumption, for suggesting the use of raking, and for providing feedback on results; Michael McDonald, D. Sunshine Hillygus, and Gustavo Novoa for pointing out sources of error in state voter files, including deadwood, and inaccuracies in the CPS (nonresponse and overreporting biases);  Eliza Lehner for discussions about BISG and election law; Cory McCartan, Kosuke Imai, and anonymous reviewers for feedback on results and exposition; and the U.S. Office of Naval Research for partial support of this work.

\bibliographystyle{abbrv}
\bibliography{refs}

\appendix
\section{The BISG independence assumption}\label{a:bisg_fl}
In this section we illustrate the impacts of BISG's conditional independence assumption
using data sets of registered voters in Florida in December 2020 and North Carolina in 
November 2010 and November 2020. In Florida and North Carolina, publicly available 
state voter files include name, address, and self-identified race/ethnicity.\footnote{In North Carolina there is a high rate of
nonresponse on self-identified race/ethnicity, whereas in Florida, nonresponse 
is relatively rare.} By using these data sets to fit BISG predictions and test their accuracy, 
we isolate the impact of BISG's conditional independence assumption.

We first describe BISG's conditional independence assumption. 
Given an individual's surname, $s$, and geolocation, $g$, the BISG prediction of 
race/ethnicity is 
\begin{align}\label{bisg}
\prob(r | g, s) 
= \frac{\prob(g | r) \, \prob(r | s)}{\sum_{i}\prob(g | r_i) \, \prob(r_i | s)}
= \frac{\prob(r | g) \, \prob(r | s) \ \prob(r)}{\sum_{i}\prob(r_i | g) \, \prob(r_i | s) / P(r_i)}
\end{align}
where $r_1,\dots,r_m$ correspond to $m$ racial groups (usually $m=6$ in practice)
and $s, g$ signify surname and geolocation. 
This formula is exact under the assumption that surname and geolocation
are conditionally independent on race/ethnicity, that is, $s | r$ and $g | r$ are
independent. 

To isolate the effects of BISG's independence assumption, we use labeled state 
voter files to both construct the factors $\prob(r), \prob(r | g), \prob(r | s)$ in BISG and to test 
the resulting BISG predictions. Specifically, each factor is evaluated in the following way.
\begin{itemize}
\item
$\prob(r | g)$ is computed for each race/ethnicity $r$ by taking the total number of 
registered voters of race $r$ in geographic region $g$ and dividing that number
by the total number of people in $g$.   

\item
$\prob(r)$ is, for each race/ethnicity $r$, the fraction of registered voters in the 
full state who identify as race/ethnicity $r$. 

\item 
$\prob(r | s)$ is computed by dividing the total number of registered voters
with surname $s$ and race/ethnicity $r$ by the total number of people with 
surname $s$. 

\end{itemize}
We then compute the BISG estimate for each registered voter. We do this same 
procedure separately for Florida in 2020, North Carolina in 2020 and North
Carolina in 2010.\footnote{We remove from this analysis those with ``inactive" registrations 
and those who did not answer the race/ethnicity question.}

\begin{figure}
\centering
  \includegraphics[width=\textwidth]{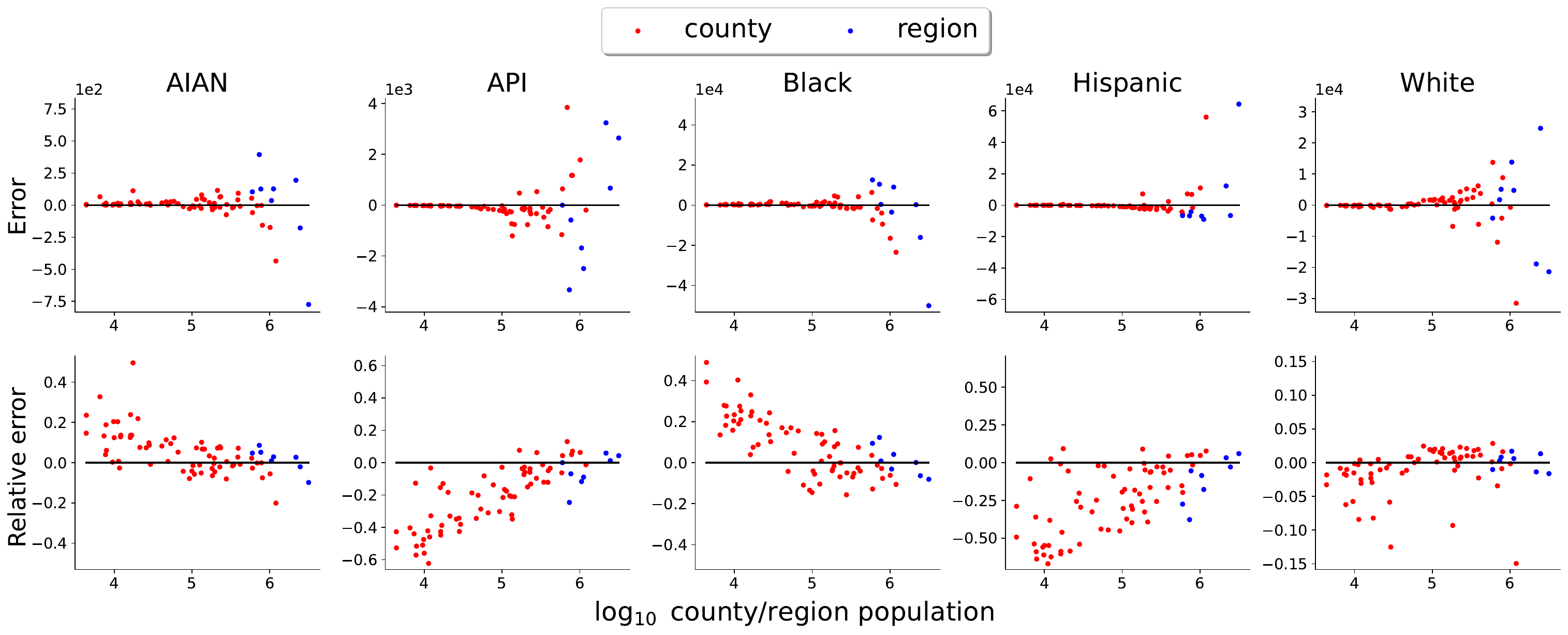}
  \caption{\em Errors in subpopulation estimation with BISG predictions 
  in each county and region in Florida in 2020. BISG was fit and tested on the
    Florida voter file.}
\label{fig:fl_vf_scatter}
\end{figure}

\begin{figure}
\centering
  \includegraphics[width=\textwidth]{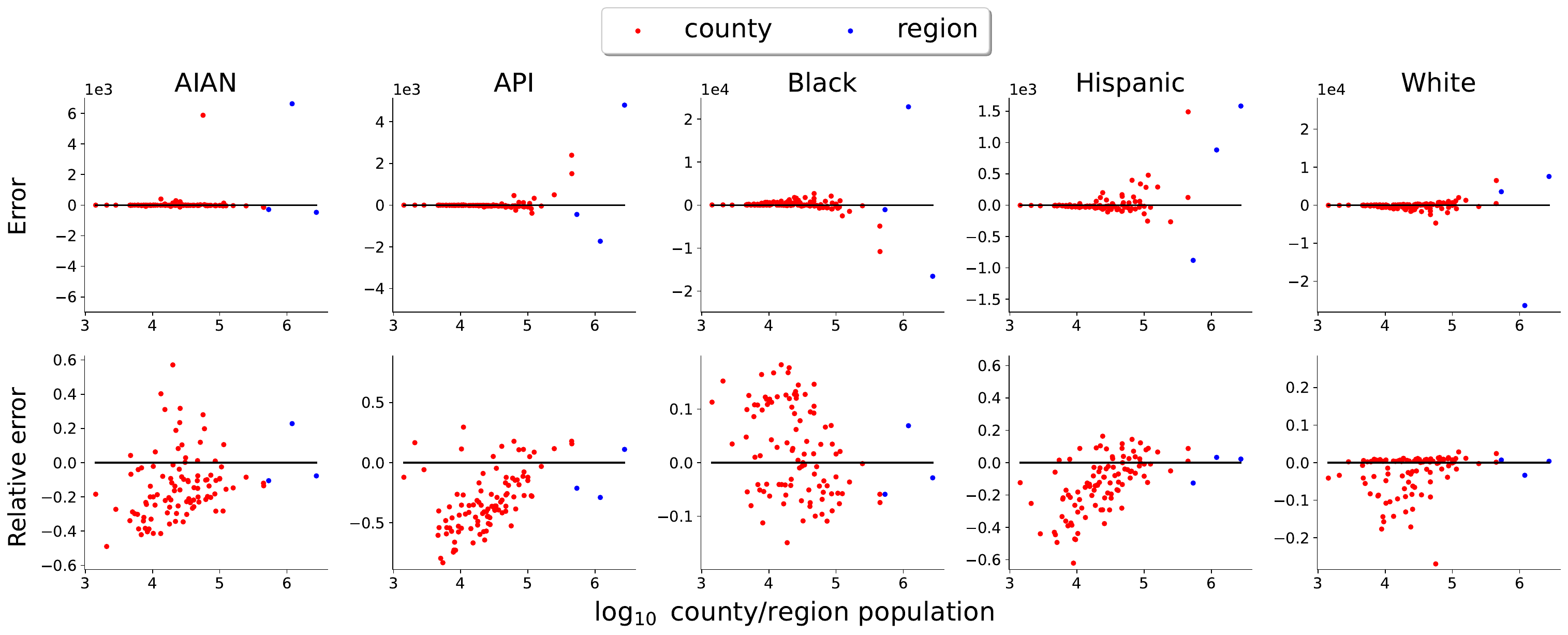}
  \caption{\em Errors in subpopulation estimation with BISG predictions 
  in each county in North Carolina in 2020. BISG was fit and tested on the
  North Carolina voter file.}
\label{fig:nc2020_vf_scatter}
\end{figure}

\begin{figure}
\centering
  \includegraphics[width=\textwidth]{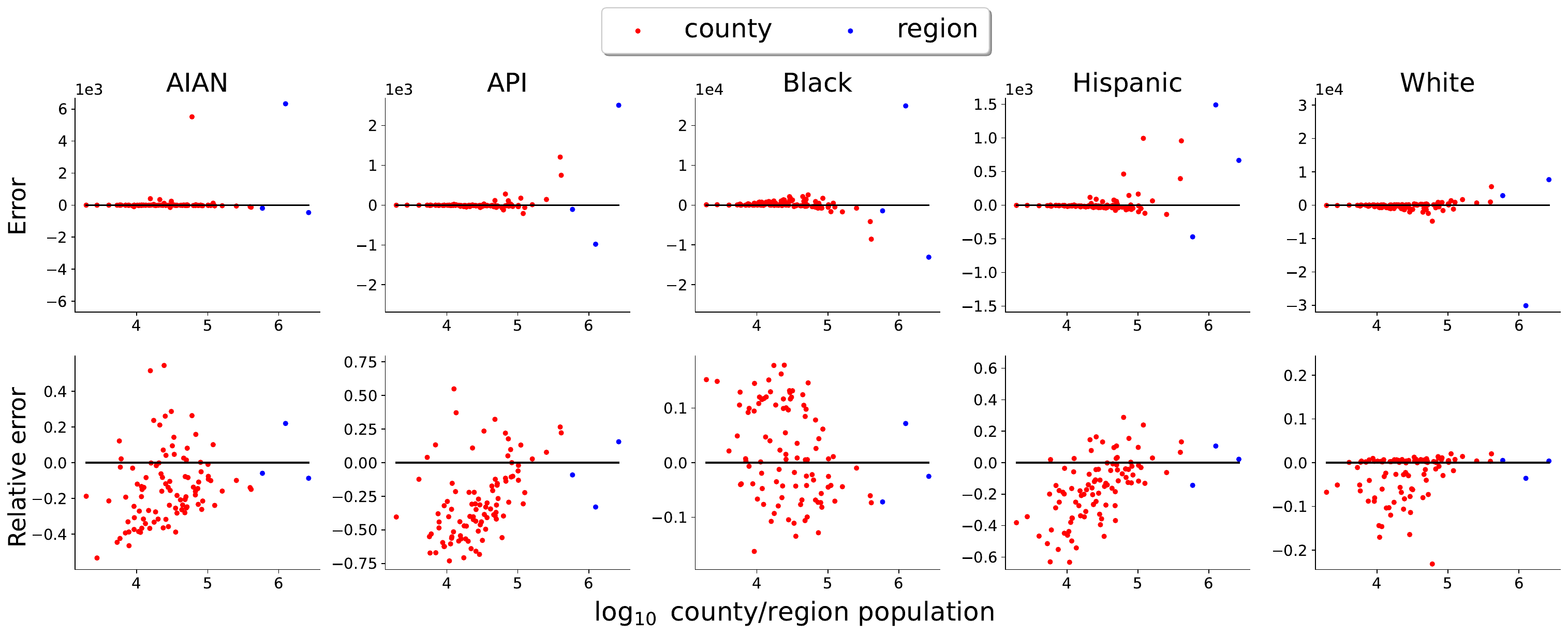}
  \caption{\em Errors in subpopulation estimation with BISG predictions 
  in each county in North Carolina in 2010. BISG was fit and tested on the
  North Carolina voter file.}
\label{fig:nc2010_vf_scatter}
\end{figure}

We use several metrics to report the accuracy of BISG fit and 
tested on Florida and North Carolina voter files. 
Recall that since each factor used in the evaluation of BISG is exact, all deviation 
from perfect accuracy is due to BISG's conditional independence assumption. 
Figures \ref{fig:fl_vf_scatter}--\ref{fig:nc2010_vf_scatter} show the absolute and relative errors of BISG subpopulation estimates in each 
county and region; see \eqref{abs_err}, \eqref{rel_err} for formal definitions of these 
metrics. 
In Table \ref{t:statewide_vf_bisg}, we report 
errors of the subpopulation estimates of BISG on the full state and in 
Table \ref{t:vf_region} we provide
region-level $\ell^1, \ell^2$ errors and negative log-likelihood (see \eqref{l1_err},
\eqref{l2_err}, \eqref{neg_ll}). Codes for generating these tables and figures as well as 
tables of county-level errors are publicly available\footnote{\url{https://github.com/pgree/raking_bisg}}. 
All metrics used for evaluating accuracy are defined in Appendix \ref{a:details}.

While statewide subpopulation estimates using BISG can be accurate,
Figures \ref{fig:fl_vf_scatter}, \ref{fig:nc2020_vf_scatter}, \ref{fig:nc2010_vf_scatter}
demonstrate that subpopulation estimation over smaller geographic regions is often 
highly inaccurate, with smaller regions and smaller subpopulations suffering from large
relative errors. In $88\%$ of counties in Florida, Hispanic 
populations were underestimated, and in $90\%$ of counties API were undercounted.
For many subpopulations, BISG errors are strongly correlated
with the size of the county population.  
County population size (and also BISG errors) 
are correlated with various features including subpopulation size, subpopulation percentage
of county population, and rurality. 
In this paper we do not investigate which of these features may cause the observed
BISG errors.

One strength of BISG is that when estimating subpopulations over the full 
state (or a reasonably large uniform sample), the errors are often small. Unfortunately, 
in practice this is not a very useful feature. Data sets are rarely uniformly sampled 
over the full population and if they are, predictions can be made using the features 
of the full population.


\begin{table}[h]
\centering
  \begin{subtable}[h]{\textwidth}
    \centering
    \resizebox{0.7\textwidth}{!}{
    \renewcommand{\arraystretch}{1.0}
    \begin{tabular}{ccccccc}

{} &    AIAN &      API &      Black &   Hispanic &      White &    Other \\
\midrule
True           &  41,132 &  246,037 &  1,762,643 &  1,933,318 &  7,799,621 &  279,677 \\
BISG           &  41,160 &  244,489 &  1,725,555 &  1,969,591 &  7,805,264 &  276,366 \\
Error          &      28 &   -1,547 &    -37,087 &     36,273 &      5,643 &   -3,310 \\
Relative Error &   0.07\% &   -0.63\% &     -2.10\% &      1.88\% &      0.07\% &   -1.18\% \\

\end{tabular}

    }
    \caption{\em Florida 2020}
    \vspace{0.5cm}
  \end{subtable}
    \begin{subtable}[h]{\textwidth}
    \centering
    \resizebox{0.7\textwidth}{!}{
    \renewcommand{\arraystretch}{1.0}
    \begin{tabular}{ccccccc}

{} &    AIAN &     API &    Black & Hispanic &      White &   Other \\
\midrule
True           &  37,843 &  51,584 &  933,877 &  104,935 &  3,277,490 &  73,432 \\
BISG           &  43,717 &  54,204 &  939,154 &  106,517 &  3,262,175 &  73,392 \\
Error          &   5,874 &   2,620 &    5,277 &    1,582 &    -15,314 &     -39 \\
Relative Error &  15.52\% &   5.08\% &    0.57\% &    1.51\% &     -0.47\% &  -0.05\% \\

\end{tabular}

    }
    \caption{\em North Carolina 2020}
    \vspace{0.5cm}
  \end{subtable}
  \begin{subtable}[h]{\textwidth}
    \centering
    \resizebox{0.7\textwidth}{!}{
    \renewcommand{\arraystretch}{1.0}
    \begin{tabular}{ccccccc}

{} &    AIAN &     API &    Black & Hispanic &      White &   Other \\
\midrule
True           &  37,337 &  20,258 &  890,308 &   48,027 &  3,434,953 &  53,215 \\
BISG           &  43,013 &  21,679 &  900,756 &   49,711 &  3,415,321 &  53,615 \\
Error          &   5,676 &   1,421 &   10,448 &    1,684 &    -19,631 &     400 \\
Relative Error &  15.20\% &   7.02\% &    1.17\% &    3.51\% &     -0.57\% &   0.75\% \\

\end{tabular}

    }
    \caption{\em North Carolina 2010}
  \end{subtable}
  \caption{\em Errors in statewide subpopulation estimation for BISG separately fit and tested 
  on each state voter file.}  
   \label{t:statewide_vf_bisg}
\end{table}


\begin{table}
\centering
  \begin{subtable}[h]{\textwidth}
    \centering
    \begin{tabular}{l | rrr | rrr | rrr } & \multicolumn{3}{
    |c|}{$\ell^1$ errors} & \multicolumn{3}{|c|}{$\ell^2$ errors} & \multicolumn{3}{|c}{negative log-likelihood} 
    \\ Region &  BISG &  $ r | g$ &  $r | s$ &  BISG &  $ r | g$ & 
     $r | s$ &  BISG &  $ r | g$ &  $r | s$  \\ \hline     Central &       0.174 &              0.813 &              0.213 &       0.107 &              0.489 &              0.134 &       0.213 &              0.455 &              0.219 \\
 Centraleast &       0.166 &              0.549 &              0.226 &       0.104 &              0.332 &              0.142 &       0.193 &              0.349 &              0.201 \\
 Centralwest &       0.152 &              0.572 &              0.219 &       0.095 &              0.344 &              0.139 &       0.194 &              0.362 &              0.204 \\
Northcentral &       0.263 &              0.614 &              0.341 &       0.169 &              0.377 &              0.218 &       0.242 &              0.382 &              0.257 \\
   Northeast &       0.188 &              0.566 &              0.298 &       0.118 &              0.341 &              0.191 &       0.236 &              0.383 &              0.254 \\
   Northwest &       0.214 &              0.405 &              0.318 &       0.135 &              0.246 &              0.202 &       0.211 &              0.292 &              0.229 \\
   Southeast &       0.176 &              0.873 &              0.272 &       0.111 &              0.526 &              0.175 &       0.224 &              0.478 &              0.245 \\
   Southwest &       0.144 &              0.529 &              0.244 &       0.091 &              0.329 &              0.156 &       0.143 &              0.302 &              0.160 \\
\hline 
     Florida &       0.175 &              0.686 &              0.253 &       0.110 &              0.414 &              0.161 &       0.209 &              0.405 &              0.223 \\

\end{tabular}

    \caption{\em Florida 2020}
    \vspace{0.5cm}
  \end{subtable}
    \begin{subtable}[h]{\textwidth}
    \centering
    \begin{tabular}{l | rrr | rrr | rrr } & \multicolumn{3}{|c|}{$\ell^1$ errors} &
 \multicolumn{3}{|c|}{$\ell^2$ errors} & \multicolumn{3}{|c}{negative log-likelihood} 
    \\ Region &  BISG &  $ r | g$ &  $r | s$ &  BISG &  $ r | g$ & 
     $r | s$ &  BISG &  $ r | g$ &  $r | s$  \\ \hline       Central &       0.199 &              0.541 &              0.264 &       0.130 &              0.336 &              0.174 &       0.206 &              0.347 &              0.217 \\
          East &       0.301 &              0.578 &              0.379 &       0.199 &              0.368 &              0.251 &       0.237 &              0.346 &              0.261 \\
          West &       0.116 &              0.199 &              0.339 &       0.076 &              0.124 &              0.226 &       0.092 &              0.139 &              0.142 \\
\hline 
North Carolina &       0.216 &              0.510 &              0.303 &       0.142 &              0.320 &              0.200 &       0.201 &              0.322 &              0.220 \\
\end{tabular}

    \caption{\em North Carolina 2020}
    \vspace{0.5cm}
  \end{subtable}
  \begin{subtable}[h]{\textwidth}
    \centering
    \begin{tabular}{l | rrr | rrr | rrr } & \multicolumn{3}{
    |c|}{$\ell^1$ errors} & \multicolumn{3}{|c|}{$\ell^2$ errors} & \multicolumn{3}{|c}{negative log-likelihood} 
    \\ Region &  BISG &  $ r | g$ &  $r | s$ &  BISG &  $ r | g$ & 
     $r | s$ &  BISG &  $ r | g$ &  $r | s$  \\ \hline       Central &       0.189 &              0.441 &              0.246 &       0.127 &              0.286 &              0.165 &       0.183 &              0.281 &              0.194 \\
          East &       0.302 &              0.544 &              0.368 &       0.203 &              0.357 &              0.248 &       0.227 &              0.316 &              0.250 \\
          West &       0.102 &              0.158 &              0.306 &       0.068 &              0.101 &              0.207 &       0.078 &              0.109 &              0.124 \\
\hline 
North Carolina &       0.209 &              0.432 &              0.288 &       0.140 &              0.281 &              0.193 &       0.182 &              0.268 &              0.200 \\

\end{tabular}

    \caption{\em North Carolina 2010}
  \end{subtable}
  \caption{\em Voter file-only fitting and testing: Accuracy of 
BISG, geolocation-only ($r|g$), and surname-only ($r|s$) predictions of race/ethnicity of registered 
voters in Florida in 2020, North Carolina in 2020, and North Carolina in 2010. 
We report $l^1$ and $l^2$ errors and negative log-likelihood for each region.}  
   \label{t:vf_region}
\end{table}


\section{BISG for registered voters}\label{a:bisg_voters}
The BISG formula \eqref{bisg} is usually calculated using census data on the 
full population.
The probability of a person belonging to a geographic region given race, $\prob(g | r)$
and the probability of race/ethnicity given surname, $\prob(r | s)$, are both obtained 
from the decennial census for the full U.S. population. 

Unfortunately, the decennial census does not provide those same distributions for 
registered voter populations in any state or the full United States. 
However, the Census Bureau does conduct the Current Population Survey (CPS) 
which contains a biannual Voter Supplement that includes questions about voter 
registration status. 
This information can be combined with the decennial census 
to obtain BISG estimates on the registered-voter population. 
We combine information from the CPS and the decennial census using an approach 
similar to that of \cite{imai2016}.  

The method of \cite{imai2016} extends BISG to incorporate demographic 
information beyond surname and geolocation (in particular
age, gender, and party registration). Their method relies on two conditional 
independence assumptions that we also use here for constructing a BISG for 
registered voters in a particular state. The assumptions are as follows:
\begin{enumerate}
\item 
For any given race/ethnicity, vote registration status and geographic region are 
independent. This conditional independence assumption implies
\begin{align}\label{vg_cond_ind}
\prob(v | r, g) = \prob(v | r)
\end{align}
where $v$ denotes vote registration status. 

\item
For any given race/ethnicity, vote registration status and surname are 
independent, implying 
\begin{align}\label{vs_cond_ind}
\prob(v | r, s) = \prob(v | r).
\end{align}

\end{enumerate}
Combining these two assumptions, we obtain a BISG formula for registered
voters. Specifically, we start by modifying the usual BISG formula 
$\prob(r | g, s) \propto \prob(r | g) \, \prob(r | s) \, / \, \prob(r)$ to condition 
on vote registration status:
\begin{align}\label{bisg_vote0}
\prob(r | g, s, v) \propto \prob(r | g, v) \, \prob(r | s, v) \, / \, \prob(r | v).
\end{align}
Bayes rule tells us 
\begin{align}\label{vote_bayes}
\prob(r | g, v) 
& \propto \prob(v | r, g) \,\, \prob(r | g) \qquad \text{and} \qquad 
\prob(r | s, v) 
 \propto \prob(v | r, s) \,\, \prob(r | s).
\end{align}
Applying the conditional independence assumptions \eqref{vg_cond_ind} and
\eqref{vs_cond_ind}, we obtain
\begin{align}\label{vote_bayes2}
\prob(r | g, v) 
& \propto \prob(v | r) \,\, \prob(r | g) \qquad \text{and} \qquad 
\prob(r | s, v) 
 \propto \prob(v | r) \,\, \prob(r | s).
\end{align}
Substituting \eqref{vote_bayes2} into \eqref{bisg_vote0}, we get a 
BISG for registered voters:
\begin{align}
\prob(r | s, g, v) \propto \prob(v | r) \, \prob(r | g) \, \prob(r | s) \, / \, \prob(r).
\end{align}
This prediction is equivalent to the usual BISG for the full population multiplied 
by $\prob(v | r)$ and appropriately normalized. 
The factor $\prob(v | r)$ is computed using the CPS estimate for
$P(r | v)$ and the census total for $P(r)$ for the above-18 population. 
This formula is equivalent to formula (7) in \cite{imai2016} where we omit the 
age and gender features $X_i$ and party registration $P_i$ is replaced with voter 
registration status.

\section{Validation via a calibration map}\label{a:calib_map}
In Section \ref{s:method} we describe a raking-based method for 
predicting race/ethnicity of registered voters. Our method fits predictions to survey 
data from the CPS on the race/ethnicity distribution of registered voters, 
and we also fit predictions to the joint distribution of 
surname and geolocation of registered voters obtained from state voter registration files. 
For validation, we test those predictions on registered voters in the voter 
files of Florida and North Carolina which contain self-identified race/ethnicity.

The accuracy of the CPS's registered voter distributions and their consistency with state voter files 
have been extensively studied in political science communities 
\cite{ansolabehere2022, mcdonald2007, fabina2022}. In \cite{mcdonald2007} McDonald 
conjectures that there are two primary sources of error in the CPS's and state voter files' 
estimates of the race/ethnicity distribution of registered voters. In voter files the primary 
source of error is likely ``deadwood,'' ``persons who are registered at but no longer live 
at an address." In the CPS, there are two main sources of error: (i) overreport bias
\cite{hammer2014,holbrook2010}, the phenomenon by which people claim to have voted, 
or be registered to vote even when they are not, and (ii) the CPS, partially to counteract 
this effect, count all ``don't know" or nonresponse as not being registered to vote. 

In this paper we do not attempt to adjust for these factors directly. Instead, 
we calibrate the CPS and voter file using two approaches. First, as described in Section 
\ref{s:results}, we subsample voter files such that the race/ethnicity distribution of the 
subsample coincides with the CPS estimate of the race/ethnicity distribution of registered 
voters in the state. That is, we ensure that the race/ethnicity distribution used for fitting 
predictions is the same as the distribution of the test set. 
In the second approach, described in this section, we construct a calibration map from 
predictions trained on the 
CPS-assumed population of registered voters to make predictions on the state voter file 
set of registered voters.\footnote{The CPS uses a different classification of race/ethnicity 
than Florida (see Appendix \ref{a:race_cat}). We use the method described in 
\cite{mcdonald2007} for converting the CPS classification to those of state voter files.}

For our calibration map, we use a $6 \times 6$ matrix (where we have $6$ races/ethnicities)
and impose two properties: (i) the matrix is ``stochastic," that is, it transforms a probability
distribution (a vector whose entries are non-negative and sum to one) to another probability 
distribution, and (ii) the matrix maps the CPS distribution to the 
voter file distribution.
Such a matrix exists for any CPS and voter file distributions. 
For example, if $u_{\text{cps}} \in \R^6$ is the probability distribution of 
the CPS and $u_{\text{vf}} \in \R^6$ is the probability distribution of the voter 
file, then the linear map $u_{\text{vf}} \mathbbm{1} $ satisfies the above conditions 
where $\mathbbm{1}$ denotes the row vector in $\R^6$ with all ones, that 
is, $\mathbbm{1} = [1,...,1]$. 

In general, a map such as $u_{\text{vf}} \mathbbm{1}$ has features that are 
undesirable for the purposes of a calibration map. Consider the simple example 
where the CPS and voter file probability distributions are identical, 
or $u_{\text{cps}} = u_{\text{vf}}$. Then $u_{vf} \mathbbm{1}$ satisfies
\begin{align}
u_{\text{vf}} \mathbbm{1} = 
\begin{bmatrix}
	\rule[.5ex]{1.3em}{0.6pt} \,\,\, u_{\text{vf}} \,\,\, \rule[.5ex]{1.3em}{0.6pt} \\
	\rule[.5ex]{1.3em}{0.6pt} \,\, u_{\text{vf}} \,\,\, \rule[.5ex]{1.3em}{0.6pt} \\
	\vdots \\
	\rule[.5ex]{1.3em}{0.6pt} \,\,\, u_{\text{vf}} \,\,\, \rule[.5ex]{1.3em}{0.6pt} \\
\end{bmatrix},
\end{align}
though if the CPS and voter file distributions are identical a more natural map would 
be the identity matrix. 
In general, the calibration map should minimize, to the extent possible, ``redistributing" 
subpopulations.\footnote{This is closely related to a set
of problems that have been formalized in the optimal transport literature. 
See for example, Section 2.3 of \cite{peyre2019}.}

To find such a matrix, we solve the following constrained convex optimization 
problem. We seek the $6 \times 6$ stochastic matrix $A$ that minimizes 
the Frobenius norm of the difference between $A$ and the identity matrix
such that $Au_{\text{cps}} = u_{\text{vf}}$. That is, we seek
\begin{equation}
\begin{aligned}
\argmin_{A} \{ \| A - I \|_{F} : 
&A_{i, j} \geq 0 \text{ for all } i, j \in \{1,...,6\}, \,
 \mathbbm{1} A  = \mathbbm{1}, \text{ and } Au_{\text{cps}} = u_{\text{vf}}\}.
\end{aligned}
\end{equation}
We solve this optimization with the Python package cvxpy
\cite{diamond2016}, \cite{agrawal2018}.

\section{Race/ethnicity classification}\label{a:race_cat}
The data sets we use in this paper---the CPS, 
the decennial census redistricting files, the surname list of 
the decennial census, and the Florida and North Carolina voter files---all 
have different categorizations of race. 
For each of these sources, we map the race/ethnicity categories provided
into six groups: American Indian and Alaskan native (AIAN), Asian 
and Pacific Islander (API), non-Hispanic Black, Hispanic, non-Hispanic White, and
other. We provide here our mapping:

\begin{itemize}
\item 
Decennial census surname list: Contains six categories---``Non-Hispanic White 
Alone, Non-Hispanic Black or African American Alone, Non-Hispanic American 
Indian and Alaska Native Alone, Non-Hispanic Asian and Native Hawaiian and 
Other Pacific Islander Alone, Non-Hispanic Two or More Races, and Hispanic or 
Latino origin." These were mapped onto our categories in the natural way with
``two or more race" being mapped onto our category ``other."

\item
Decennial census redistricting files: Respondents are asked for 
Hispanic/non-Hispanic origin and in a separate question respondents are 
asked which race they identify as among AIAN, API, Black, White, 
Other, and two or more races. To map onto our categorization, all those of 
Hispanic origin were included in Hispanic. Those who identified as two or more
races in the census were placed in the ``other" group. 

\item 
Current population survey: Like in the decennial census, all respondents are 
asked Hispanic/non-Hispanic origin and separately respondents are asked 
detailed race questions. 
For this mapping, we use the classification of \cite{mcdonald2007} : 
``All CPS respondents reporting Hispanic ethnicity are scored as Hispanic. 
All non-Hispanics reporting a single race only are reported as that race. 
Asian and Hawaiian-Pacific Islander are grouped into an Asian category. 
For multiple-race categories, non-Hispanics reporting Black in any
other combination are scored as Black. Among the remainder, non-Hispanics 
reporting Asian or Hawaiian-Pacific Islander in combination with any other 
remaining race are identified as Asian. Those remaining are classified as other."

\item 
Florida voter file: 
Includes race/ethnicity groups: ``American Indian/ Alaskan Native" (AIAN), 
``Asian/Pacific Islander (API),'' ``Black, {\it not of} Hispanic Origin," 
``Hispanic," ``White, {\it not of} Hispanic Origin," ``Multi-racial," ``Other."
We included all ``multi-racial" in our ``other" bucket. 

\item North Carolina voter file: Includes separate fields for Hispanic 
origin and race. All Hispanics are included in the ``Hispanic" group. Among the 
remaining, classifications were made in the natural way and "multi-racial" were 
included in "other."

\end{itemize}

\section{Detailed results}\label{a:details}
We use several error metrics for validation in this paper. Here, we provide formulas for these
metrics. As in Section \ref{s:method}, we denote by $x_{sgr}$ the correct values
of the surname by geolocation by race/ethnicity contingency table and we denote
by $m_{sgr}$ the corresponding predictions. We denote by a subscript ``+" summing 
over an index. So, $x_{+gr}$ and $m_{+gr}$ denote summing over the surname index 
for the correct and predicted values respectively. \\

\noindent {\bf Subpopulation estimation} 
\begin{itemize}
\item
Absolute error: 
\begin{align}\label{abs_err}
m_{+gr} - x_{+gr}
\end{align}
for $g = 1,\dots,n_{g}$ and $r = 1,\dots,6$. 

\item
Relative error: 
\begin{align}\label{rel_err}
\frac{m_{+gr} - x_{+gr}}{x_{+gr}}
\end{align}
for $g = 1,\dots,n_{g}$ and $r = 1,\dots,6$. 

\item
Mean absolute deviation:
\begin{align}\label{mad}
\sum_{g=1}^{n_g} | m_{+gr} - x_{+gr} |
\end{align}
for $r = 1,\dots,6$. 

\item
Average errors:
\begin{align}\label{avg_err}
m_{++r} - x_{++r}
\end{align}
for $r = 1,\dots,6$. 

\end{itemize}
\vspace{0.5cm}

\noindent {\bf $\ell^1, \ell^2$ and negative log-likelihood}
\begin{itemize}
\item
$\ell^1$ error: 
\begin{align}\label{l1_err}
\frac{1}{x_{+g+}} \sum_{s=1}^{n_s} \sum_{r=1}^{6} | m_{sgr} - x_{sgr} |
\end{align}
for $g = 1,\dots,n_{g}$.

\item
$\ell^2$ error: 
\begin{align}\label{l2_err}
\frac{1}{x_{+g+}} \sum_{s=1}^{n_s} \bigg( \sum_{r=1}^{6}  (m_{sgr} - x_{sgr})^2\bigg)^{1/2}
\end{align}
for $g = 1,\dots,n_{g}$.

\item
Negative log-likelihood:
\begin{align}\label{neg_ll}
-\frac{1}{x_{+g+}} \sum_{s=1}^{n_s} \sum_{r=1}^{6} x_{sgr}\log(m_{sgr})
\end{align}

\end{itemize}

All raking and BISG predictions of this section are computed using the 
methods described in Section \ref{s:results} and details of the BISG implementation
are provided in Appendix \ref{a:bisg_voters}.
Raking predictions are computed by raking to CPS race/ethnicity margins and surname 
by geolocation margins of state voter files. 

\begin{figure}
\centering
\begin{subfigure}[b]{0.99\linewidth}
\centering
  \includegraphics[height=8.2cm, keepaspectratio]{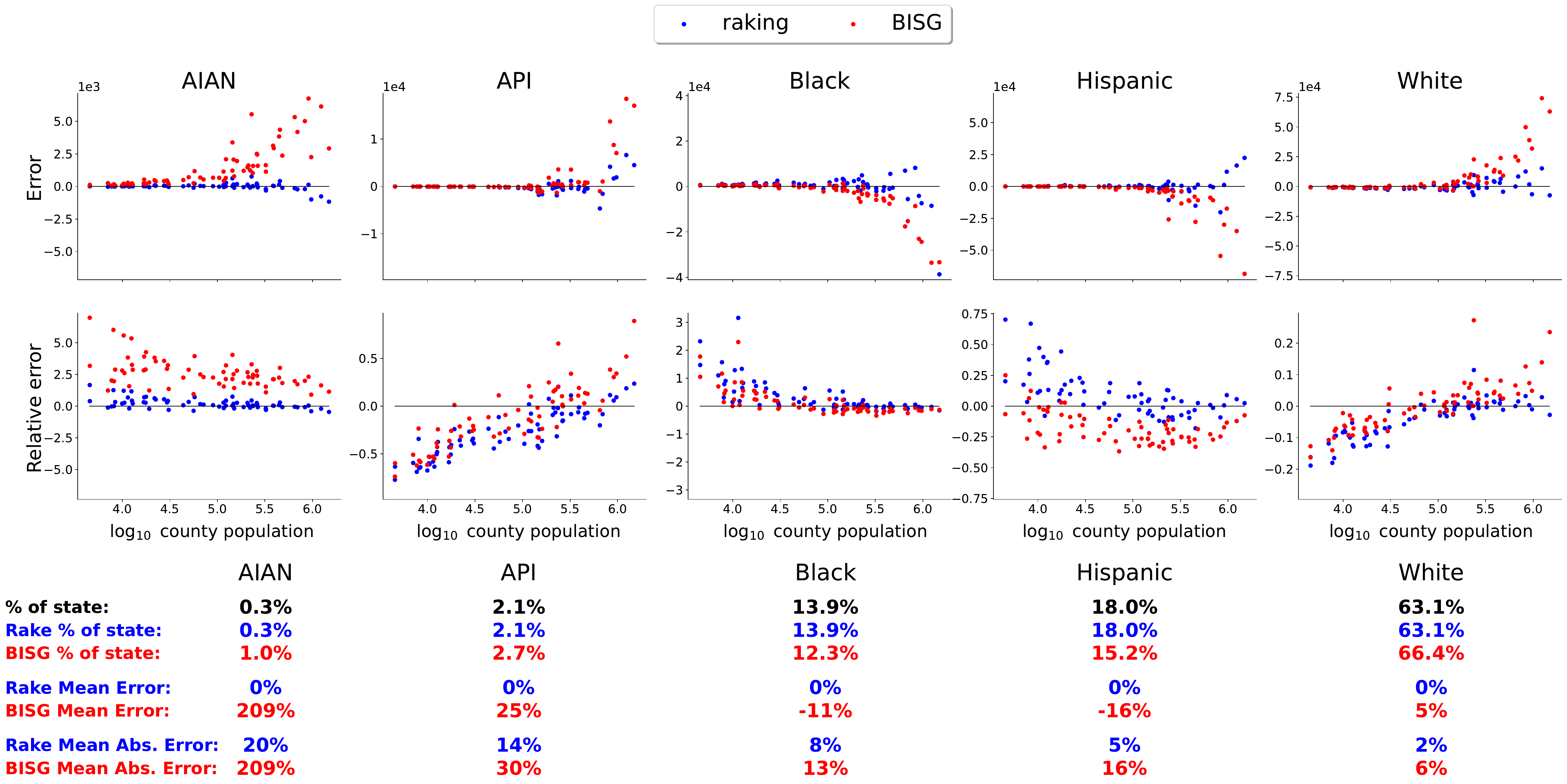}
\end{subfigure}
  \caption{\em Errors in subpopulation estimation for registered voters in each county in Florida in 2020 for two methods, BISG and raking, using a calibration map. Each dot in the scatterplots represents error in one county with BISG or raking. Dots above the horizontal line correspond to overestimating the size of a subpopulation size in one county. Dots below the horizontal line indicate underestimating.}
    \label{fig:fl_scatter_bar}
\end{figure}

\begin{figure}
\centering
\begin{subfigure}[b]{0.99\linewidth}
\centering
  \includegraphics[height=8.2cm, keepaspectratio]{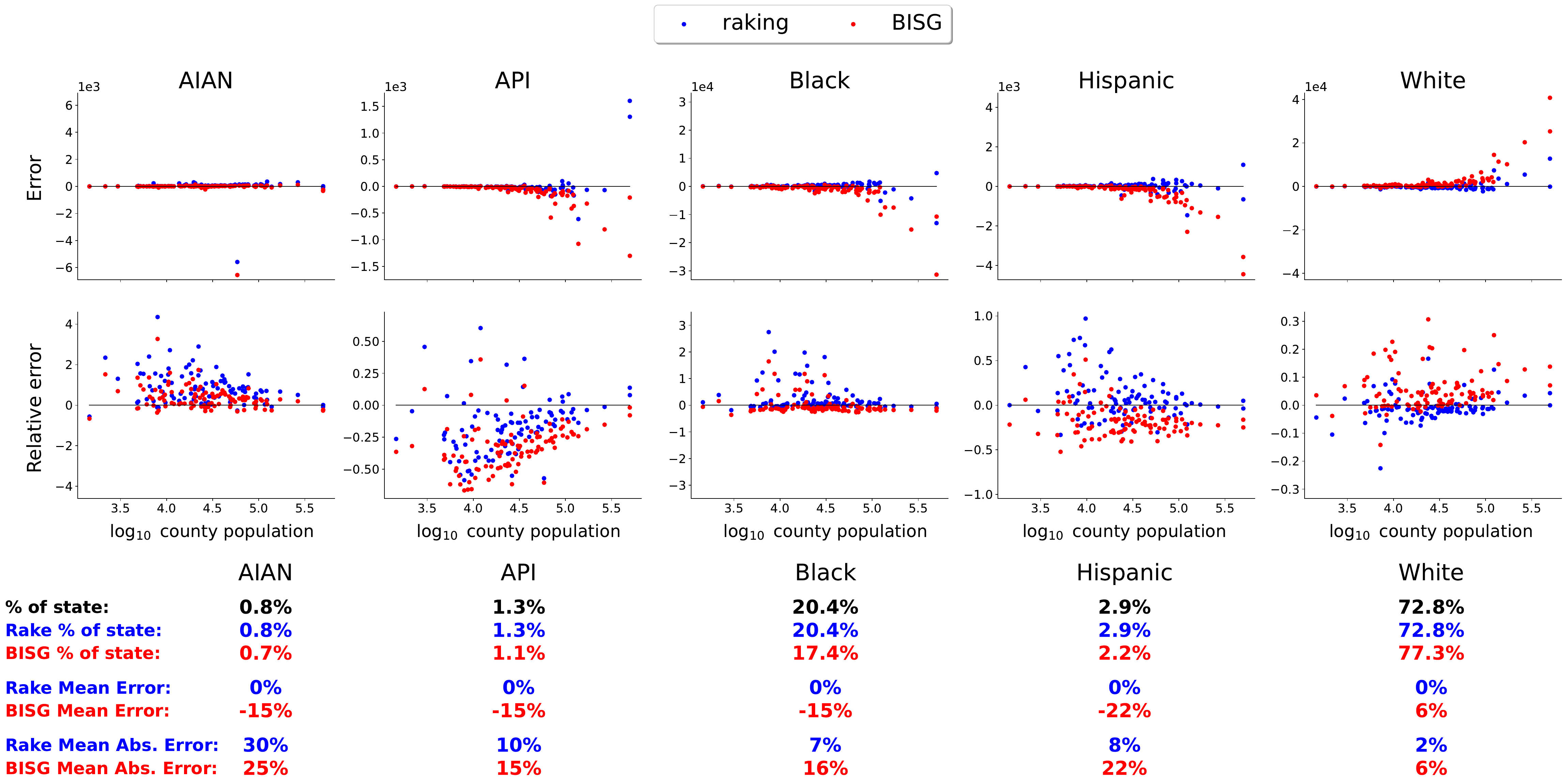}
\end{subfigure}
  \caption{\em Errors in subpopulation estimation for registered voters in each county in North Carolina in 2020 for two methods, BISG and raking, using a calibration map. Each dot in the scatterplots represents error of one subpopulation size in one county with BISG or raking. Dots above the horizontal line correspond to overestimating the size of a subpopulation size in one county. Dots below the horizontal line indicate underestimating.}
    \label{fig:nc2020_scatter_bar}
\end{figure}

\begin{figure}
\centering
\begin{subfigure}[b]{0.99\linewidth}
\centering
  \includegraphics[height=8.2cm, keepaspectratio]{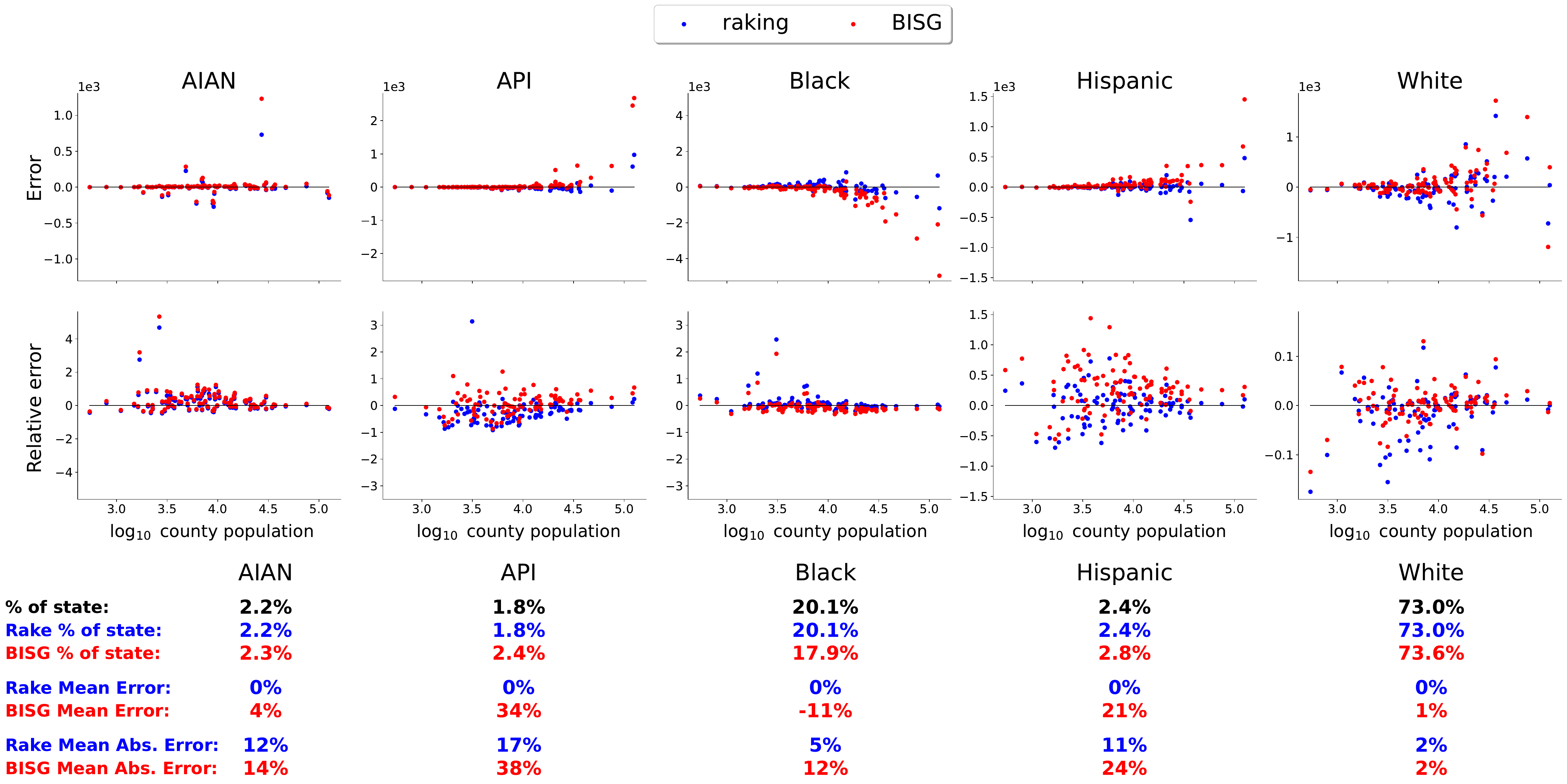}
\end{subfigure}
  \caption{\em Errors in subpopulation estimation for subsampled registered voters in each county in North Carolina in 2010 for two methods, BISG and raking. Each dot in the scatterplots represents error in one county with BISG or raking. Dots above the horizontal line correspond to overestimating the size of a subpopulation size in one county. Dots below the horizontal line indicate underestimating.}
    \label{fig:nc2010_scatter_bar_sub}
\end{figure}

\begin{figure}
\centering
\begin{subfigure}[b]{0.99\linewidth}
\centering
  \includegraphics[height=8.2cm, keepaspectratio]{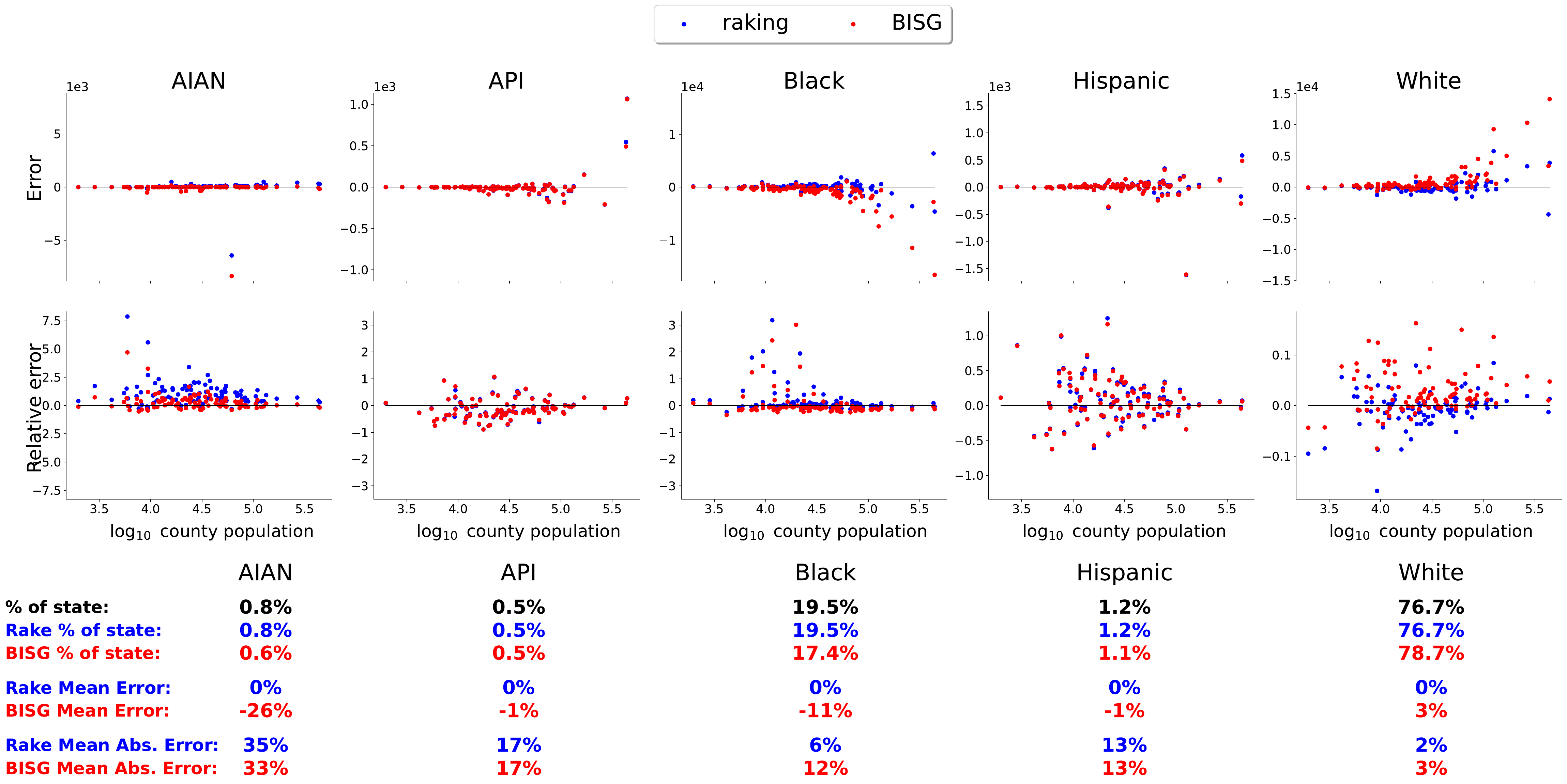}
\end{subfigure}
  \caption{\em Errors in subpopulation estimation for registered voters in each county in North Carolina in 2010 for two methods, BISG and raking, using a calibration map. Each dot in the scatterplots represents error of one subpopulation size in one county with BISG or raking. Dots above the horizontal line correspond to overestimating the size of a subpopulation size in one county. Dots below the horizontal line indicate underestimating.}
    \label{fig:nc2010_scatter_bar}
\end{figure}

We illustrate the accuracy of raking and BISG  for both approaches to validation, subsampling 
and the calibration map (see Appendix \ref{a:calib_map}) in the following figures and tables. 
We include subpopulation estimation accuracy in Florida in 2020 with subsampling in 
Figure \ref{fig:p1} and calibration map in Figure \ref{fig:fl_scatter_bar}. 
Those same metrics are reported in North Carolina in 2020 
(Figures \ref{fig:nc2020_scatter_bar_sub} and \ref{fig:nc2020_scatter_bar}), and 
in North Carolina in 2010 (Figures \ref{fig:nc2010_scatter_bar_sub} and \ref{fig:nc2010_scatter_bar}). 
In each of the above figures, the scatterplots provide absolute and relative errors \eqref{abs_err} 
and \eqref{rel_err} and each dot corresponds to one subpopulation in one county. 
The bar plots of the above figures report mean absolute deviation \eqref{mad} and 
average error \eqref{avg_err}.

\begin{table}
\centering
  \begin{subtable}[h]{\textwidth}
    \centering
    \resizebox{0.7\textwidth}{!}{
    \resizebox{0.8\textwidth}{!}{ \begin{tabular}{l | cc | cc | cc } & \multicolumn{2}{|c|}{
    $\ell^1$ errors} & \multicolumn{2}{|c|}{$\ell^2$ errors} & \multicolumn{2}{|c}{negative log-likelihood} \\ 
    Region &  raking &  BISG &  raking &  BISG &  raking &  BISG \\ \hline     Central &    0.362 &          0.365 &    0.227 &          0.227 &    0.271 &          0.285 \\
 Centraleast &    0.260 &          0.261 &    0.165 &          0.165 &    0.219 &          0.226 \\
 Centralwest &    0.266 &          0.270 &    0.168 &          0.169 &    0.228 &          0.235 \\
Northcentral &    0.352 &          0.358 &    0.230 &          0.232 &    0.255 &          0.260 \\
   Northeast &    0.286 &          0.293 &    0.181 &          0.184 &    0.260 &          0.265 \\
   Northwest &    0.269 &          0.274 &    0.172 &          0.172 &    0.220 &          0.224 \\
   Southeast &    0.393 &          0.402 &    0.245 &          0.249 &    0.281 &          0.286 \\
   Southwest &    0.257 &          0.234 &    0.165 &          0.149 &    0.179 &          0.181 \\
\hline 
     Florida &    0.323 &          0.327 &    0.204 &          0.205 &    0.250 &          0.257 \\

\end{tabular}
} 
    } 
    \caption{\em Florida 2020, subsampling}
    \vspace{0.5cm}
  \end{subtable}
    \begin{subtable}[h]{\textwidth}
    \centering
    \resizebox{0.7\textwidth}{!}{
    \resizebox{0.8\textwidth}{!}{ \begin{tabular}{l | cc | cc | cc } & \multicolumn{2}{|c|}{
    $\ell^1$ errors} & \multicolumn{2}{|c|}{$\ell^2$ errors} & \multicolumn{2}{|c}{negative log-likelihood} \\ 
    Region &  raking &  BISG &  raking &  BISG &  raking &  BISG \\ \hline     Central &    0.381 &          0.387 &    0.231 &          0.234 &    0.305 &          0.319 \\
 Centraleast &    0.281 &          0.283 &    0.173 &          0.173 &    0.250 &          0.256 \\
 Centralwest &    0.286 &          0.289 &    0.174 &          0.175 &    0.257 &          0.264 \\
Northcentral &    0.383 &          0.384 &    0.242 &          0.240 &    0.293 &          0.297 \\
   Northeast &    0.305 &          0.311 &    0.187 &          0.190 &    0.291 &          0.295 \\
   Northwest &    0.295 &          0.293 &    0.183 &          0.178 &    0.246 &          0.249 \\
   Southeast &    0.404 &          0.415 &    0.243 &          0.249 &    0.317 &          0.323 \\
   Southwest &    0.275 &          0.257 &    0.170 &          0.158 &    0.202 &          0.205 \\
\hline 
     Florida &    0.341 &          0.345 &    0.208 &          0.210 &    0.282 &          0.289 \\

\end{tabular}
} 
    } 
    \caption{\em Florida 2020, calibration map}
  \end{subtable}
  \caption{\em Florida 2020 validation via subsampling and calibration map:
  $\ell^1, \ell^2$ errors and negative log-likelihood in each region. }  
   \label{t:fl_validation_region}
\end{table}

\begin{table}
\centering
  \begin{subtable}[h]{\textwidth}
    \centering
    \resizebox{0.7\textwidth}{!}{
    \resizebox{0.8\textwidth}{!}{ \begin{tabular}{l | cc | cc | cc } & \multicolumn{2}{|c|}{
    $\ell^1$ errors} & \multicolumn{2}{|c|}{$\ell^2$ errors} & \multicolumn{2}{|c}{negative log-likelihood} \\ 
    Region &  raking &  BISG &  raking &  BISG &  raking &  BISG \\ \hline       Central &    0.379 &          0.381 &    0.239 &          0.242 &    0.287 &          0.291 \\
          East &    0.461 &          0.458 &    0.297 &          0.296 &    0.308 &          0.309 \\
          West &    0.218 &          0.205 &    0.139 &          0.132 &    0.141 &          0.140 \\
\hline 
North Carolina &    0.383 &          0.382 &    0.244 &          0.244 &    0.276 &          0.279 \\

\end{tabular}
} 
    } 
    \caption{\em North Carolina 2020, subsampling}
    \vspace{0.5cm}
  \end{subtable}
    \begin{subtable}[h]{\textwidth}
    \centering
    \resizebox{0.7\textwidth}{!}{
    \resizebox{0.8\textwidth}{!}{ \begin{tabular}{l | cc | cc | cc } & \multicolumn{2}{|c|}{
    $\ell^1$ errors} & \multicolumn{2}{|c|}{$\ell^2$ errors} & \multicolumn{2}{|c}{negative log-likelihood} \\ 
    Region &  raking &  BISG &  raking &  BISG &  raking &  BISG \\ \hline       Central &    0.350 &          0.340 &    0.218 &          0.217 &    0.265 &          0.268 \\
          East &    0.436 &          0.424 &    0.278 &          0.275 &    0.287 &          0.290 \\
          West &    0.189 &          0.159 &    0.116 &          0.100 &    0.117 &          0.115 \\
\hline 
North Carolina &    0.354 &          0.341 &    0.222 &          0.218 &    0.254 &          0.256 \\

\end{tabular}
} 
    } 
    \caption{\em North Carolina 2020, calibration map}
  \end{subtable}
  \caption{\em North Carolina 2020 validation via subsampling and calibration map: 
  $\ell^1, \ell^2$ errors and negative log-likelihood in each region. }  
   \label{t:nc2020_validation_region}
\end{table}

\begin{table}
\centering
  \begin{subtable}[h]{\textwidth}
    \centering
    \resizebox{0.7\textwidth}{!}{
    \resizebox{0.8\textwidth}{!}{ \begin{tabular}{l | cc | cc | cc } & \multicolumn{2}{|c|}{
    $\ell^1$ errors} & \multicolumn{2}{|c|}{$\ell^2$ errors} & \multicolumn{2}{|c}{negative log-likelihood} \\ 
    Region &  raking &  BISG &  raking &  BISG &  raking &  BISG \\ \hline       Central &    0.323 &          0.336 &    0.211 &          0.215 &    0.240 &          0.243 \\
          East &    0.445 &          0.452 &    0.293 &          0.293 &    0.285 &          0.286 \\
          West &    0.173 &          0.178 &    0.113 &          0.115 &    0.113 &          0.113 \\
\hline 
North Carolina &    0.339 &          0.349 &    0.222 &          0.224 &    0.236 &          0.239 \\

\end{tabular}
} 
    } 
    \caption{\em North Carolina 2010, subsampling}
    \vspace{0.5cm}
  \end{subtable}
    \begin{subtable}[h]{\textwidth}
    \centering
    \resizebox{0.7\textwidth}{!}{
    \resizebox{0.8\textwidth}{!}{ \begin{tabular}{l | cc | cc | cc } & \multicolumn{2}{|c|}{
    $\ell^1$ errors} & \multicolumn{2}{|c|}{$\ell^2$ errors} & \multicolumn{2}{|c}{negative log-likelihood} \\ 
    Region &  raking &  BISG &  raking &  BISG &  raking &  BISG \\ \hline       Central &    0.288 &          0.288 &    0.187 &          0.186 &    0.223 &          0.224 \\
          East &    0.412 &          0.409 &    0.269 &          0.267 &    0.268 &          0.269 \\
          West &    0.147 &          0.139 &    0.093 &          0.088 &    0.095 &          0.095 \\
\hline 
North Carolina &    0.304 &          0.302 &    0.197 &          0.196 &    0.219 &          0.220 \\

\end{tabular}
} 
    } 
    \caption{\em North Carolina 2010, calibration map}
  \end{subtable}
  \caption{\em North Carolina 2010 validation via subsampling and calibration map:
   $\ell^1, \ell^2$ errors and negative log-likelihood in each region. }  
   \label{t:nc2010_validation_region}
\end{table}

We also include $\ell^1, \ell^2$ errors and negative log-likelihood (see \eqref{l1_err}, \eqref{l2_err}, and \eqref{neg_ll}) by region in 
Florida in 2020 (Table \ref{t:fl_validation_region}), 
North Carolina in 2020 (Table \ref{t:nc2020_validation_region}), and 
North Carolina in 2010 (Table \ref{t:nc2010_validation_region}). 

\begin{table}
\centering
  \begin{subtable}[h]{\textwidth}
    \centering
    \resizebox{0.6\textwidth}{!}{
    \begin{tabular}{lrrrrrr}

        Population &  AIAN &   API &  Black &  Hispanic &  White &  Other \\
\midrule
   USA census 2020 & 0.007 & 0.061 &  0.121 &     0.187 &  0.578 &  0.046 \\
    FL 2020 census & 0.002 & 0.030 &  0.145 &     0.265 &  0.515 &  0.043 \\
FL 2020 18+ census & 0.002 & 0.030 &  0.135 &     0.250 &  0.547 &  0.036 \\
FL 2020 CPS voters & 0.004 & 0.028 &  0.141 &     0.190 &  0.636 &  0.001 \\
FL 2020 voter file & 0.003 & 0.021 &  0.139 &     0.180 &  0.631 &  0.025 \\

\end{tabular}

    } 
    \caption{\em Florida 2020}
    \vspace{0.5cm}
  \end{subtable}
  \begin{subtable}[h]{\textwidth}
    \centering
    \resizebox{0.6\textwidth}{!}{
    \begin{tabular}{lrrrrrr}

        Population &  AIAN &   API &  Black &  Hispanic &  White &  Other \\
\midrule
   USA census 2020 & 0.007 & 0.061 &  0.121 &     0.187 &  0.578 &  0.046 \\
    NC 2020 census & 0.010 & 0.033 &  0.202 &     0.107 &  0.605 &  0.043 \\
NC 2020 18+ census & 0.009 & 0.032 &  0.199 &     0.089 &  0.636 &  0.034 \\
NC 2020 CPS voters & 0.019 & 0.031 &  0.227 &     0.052 &  0.662 &  0.009 \\
NC 2020 voter file & 0.008 & 0.013 &  0.204 &     0.029 &  0.728 &  0.018 \\

\end{tabular}

    } 
    \caption{\em North Carolina 2020}
    \vspace{0.5cm}
  \end{subtable}
    \begin{subtable}[h]{\textwidth}
    \centering
    \resizebox{0.6\textwidth}{!}{
    \begin{tabular}{lrrrrrr}

        Population &  AIAN &   API &  Black &  Hispanic &  White &  Other \\
\midrule
   USA census 2010 & 0.007 & 0.048 &  0.122 &     0.163 &  0.637 &  0.021 \\
    NC 2010 census & 0.011 & 0.022 &  0.212 &     0.084 &  0.653 &  0.018 \\
NC 2010 18+ census & 0.011 & 0.022 &  0.204 &     0.068 &  0.684 &  0.011 \\
NC 2010 CPS voters & 0.022 & 0.018 &  0.201 &     0.024 &  0.730 &  0.005 \\
NC 2010 voter file & 0.008 & 0.005 &  0.195 &     0.012 &  0.767 &  0.013 \\

\end{tabular}

    } 
    \caption{\em North Carolina 2010}
  \end{subtable}
  \caption{\em Various race/ethnicity distributions in Florida and North Carolina.}  
   \label{t:summary_table}
\end{table}

Table \ref{t:summary_table} provides various relevant race/ethnicity distributions including 
those of state voter files, the CPS, and the decennial census.

\begin{figure}
\centering
\begin{subfigure}{0.99\linewidth}
\centering
  \includegraphics[width=0.6\textwidth, keepaspectratio]{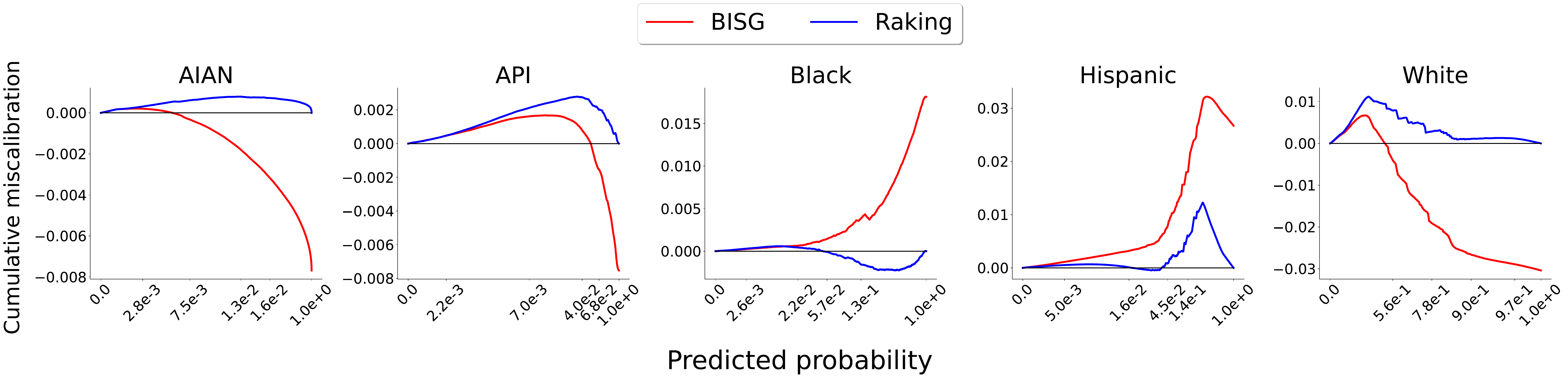}
  \caption{\em 2020 Florida, subampled}
\end{subfigure}
\hspace{0.5cm}
\begin{subfigure}{0.99\linewidth}
\centering
  \includegraphics[width=0.6\textwidth, keepaspectratio]{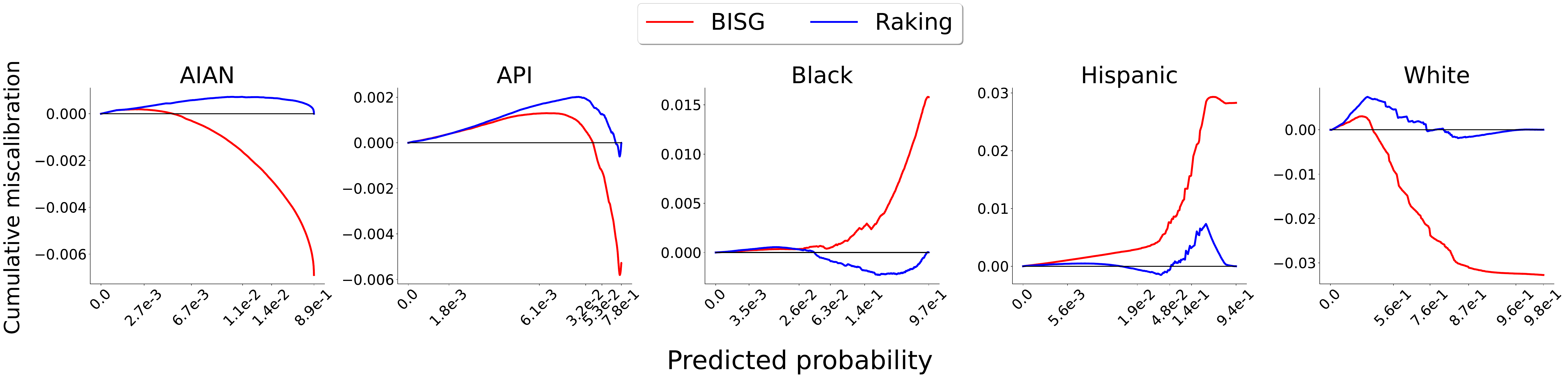}
  \caption{\em 2020 Florida, calibration map}
\end{subfigure}\par\bigskip 
\begin{subfigure}{0.99\linewidth}
\centering
  \includegraphics[width=0.6\textwidth, keepaspectratio]{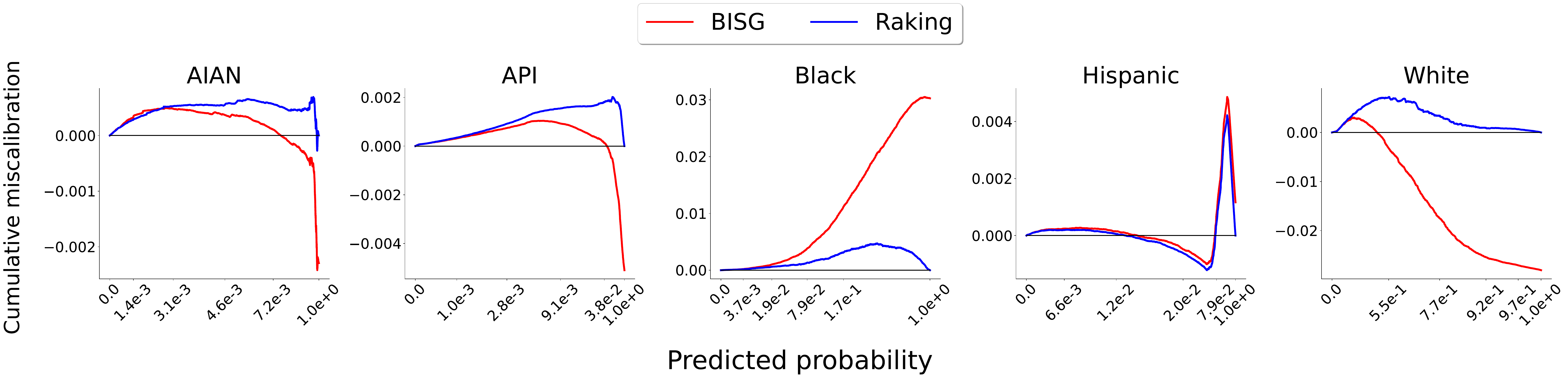}
  \caption{\em 2020 North Carolina, subampled}
\end{subfigure}
\hspace{0.5cm}
\begin{subfigure}{0.99\linewidth}
\centering
  \includegraphics[width=0.6\textwidth, keepaspectratio]{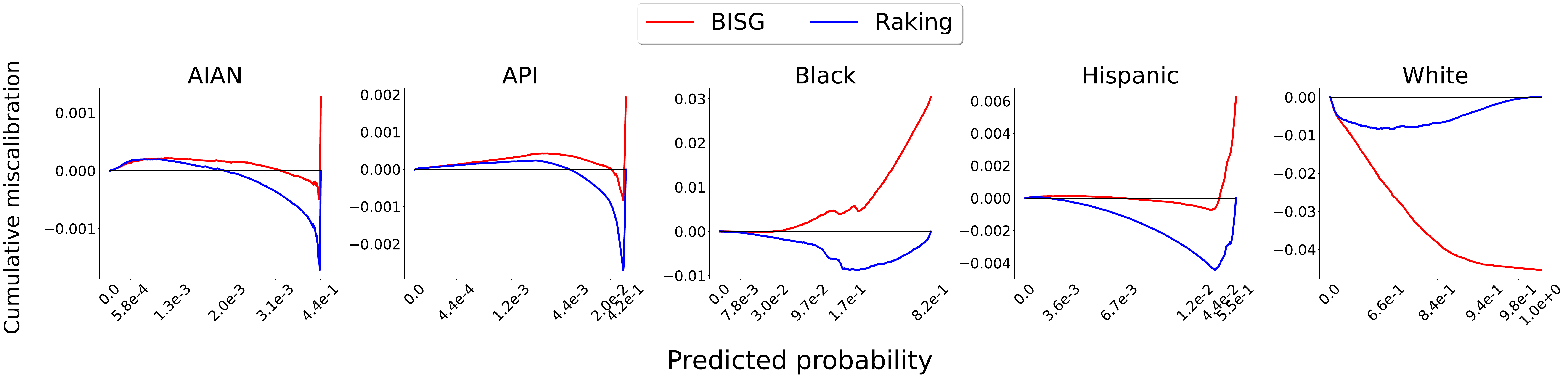}
  \caption{\em 2020 North Carolina, calibration map}
\end{subfigure}\par\bigskip 
\begin{subfigure}{0.99\linewidth}
\centering
  \includegraphics[width=0.6\textwidth, keepaspectratio]{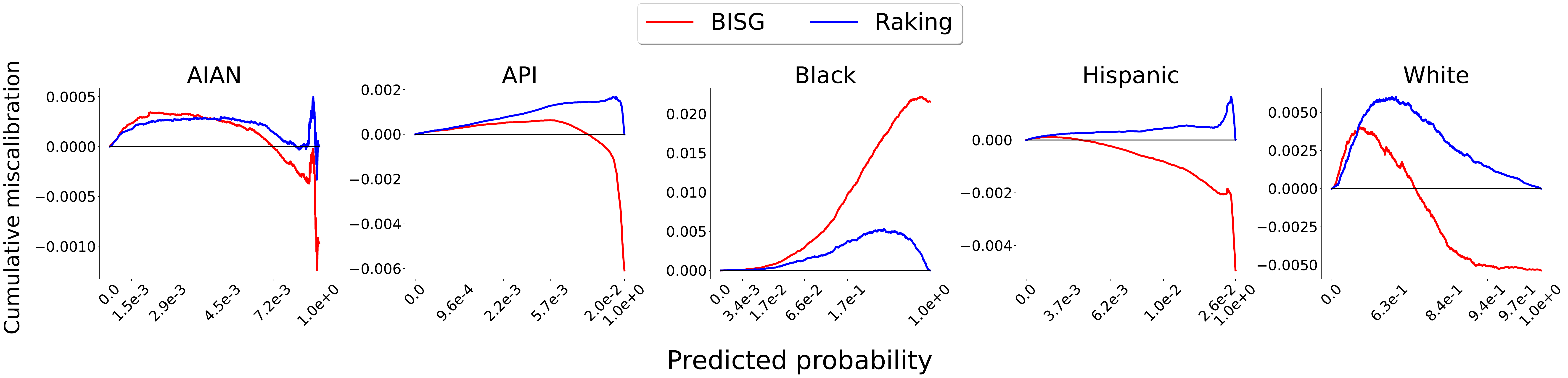}
  \caption{\em 2010 North Carolina, subampled}
\end{subfigure}
\hspace{0.5cm}
\begin{subfigure}{0.99\linewidth}
\centering
  \includegraphics[width=0.6\textwidth, keepaspectratio]{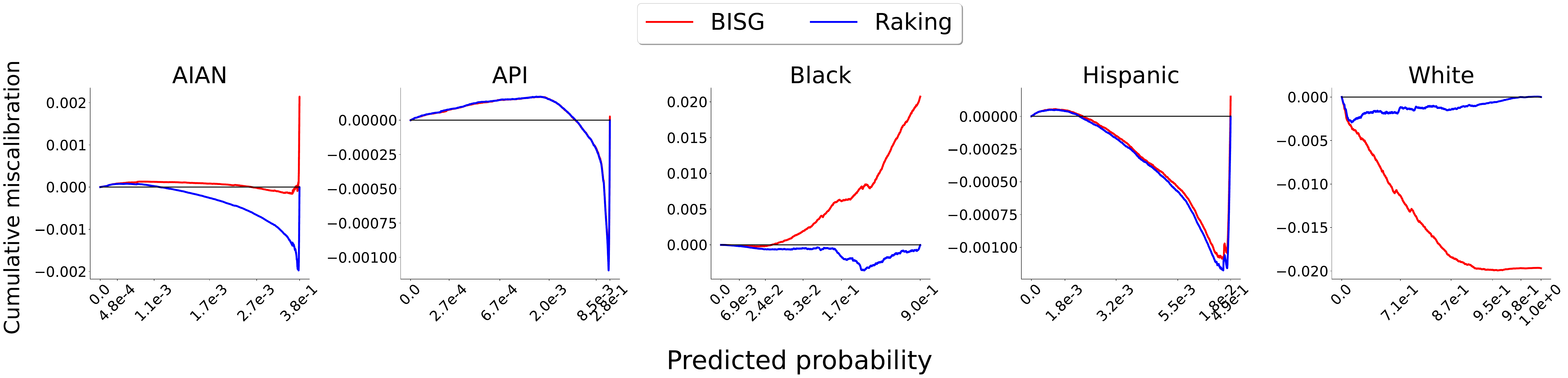}
  \caption{\em 2010 North Carolina, calibration map}
\end{subfigure}\par\bigskip 
\caption{\em Cumulative miscalibration of BISG and raking on 
  registered voters in Florida and North Carolina. The vertical axis represents cumulative deviation from
  perfect calibration. The slope of a line connecting any two
  points on the curve is the average miscalibration for the predicted probabilities between
  those two points. Flatter curves near zero represent better calibration. 
  The ticks on the horizontal axis are quintiles of BISG 
  predictions. Non-uniform spacing between quintiles is due to weighting more 
  frequently-appearing (surname, geolocation) pairs in estimating miscalibration.}
    \label{fig:calib_grid}
\end{figure}

\begin{table}
\centering
    \begin{subtable}[h]{\linewidth}
        \centering
        \begin{tabular}{lrrrrr}

{} &   AIAN &    API &  Black &  Hispanic &  White \\
\midrule
FL 2020 BISG   & 0.0079 & 0.0092 & 0.0182 &    0.0322 & 0.0371 \\
FL 2020 raking & 0.0008 & 0.0028 & 0.0029 &    0.0128 & 0.0112 \\

\end{tabular}

 	\caption{\em Florida 2020, subsampled}
        \vspace{0.3cm}
    \end{subtable}
    \begin{subtable}[h]{\textwidth}
        \centering
        \begin{tabular}{lrrrrr}

{} &   AIAN &    API &  Black &  Hispanic &  White \\
\midrule
FL 2020 BISG   & 0.0071 & 0.0071 & 0.0158 &    0.0293 & 0.0358 \\
FL 2020 raking & 0.0007 & 0.0026 & 0.0028 &    0.0088 & 0.0093 \\

\end{tabular}

 	\caption{\em Florida 2020, calibration map}
        \vspace{0.3cm}
    \end{subtable}
    \begin{subtable}[h]{\textwidth}
        \centering
        \begin{tabular}{lrrrrr}

{} &   AIAN &    API &  Black &  Hispanic &  White \\
\midrule
NC 2020 BISG   & 0.0029 & 0.0061 & 0.0305 &    0.0059 & 0.0311 \\
NC 2020 raking & 0.0010 & 0.0020 & 0.0047 &    0.0054 & 0.0072 \\

\end{tabular}

 	\caption{\em North Carolina 2020, subsampled}
         \vspace{0.3cm}
     \end{subtable}
    \begin{subtable}[h]{\textwidth}
        \centering
        \begin{tabular}{lrrrrr}

{} &   AIAN &    API &  Black &  Hispanic &  White \\
\midrule
NC 2020 BISG   & 0.0018 & 0.0028 & 0.0306 &    0.0070 & 0.0454 \\
NC 2020 raking & 0.0019 & 0.0029 & 0.0088 &    0.0045 & 0.0085 \\

\end{tabular}

 	\caption{\em North Carolina 2020, calibration map}
          \label{t:kuiper_nc2020}
         \vspace{0.3cm}
     \end{subtable}
     \begin{subtable}[h]{\textwidth}
        \centering
        \begin{tabular}{lrrrrr}

{} &   AIAN &    API &  Black &  Hispanic &  White \\
\midrule
NC 2010 BISG   & 0.0016 & 0.0067 & 0.0222 &    0.0051 & 0.0094 \\
NC 2010 raking & 0.0008 & 0.0017 & 0.0053 &    0.0016 & 0.0060 \\

\end{tabular}

 	\caption{\em North Carolina 2010, subsampled}
         \vspace{0.3cm}
     \end{subtable}
   \begin{subtable}[h]{\textwidth}
        \centering
        \begin{tabular}{lrrrrr}

{} &   AIAN &    API &  Black &  Hispanic &  White \\
\midrule
NC 2010 BISG   & 0.0023 & 0.0013 & 0.0210 &    0.0012 & 0.0199 \\
NC 2010 raking & 0.0021 & 0.0013 & 0.0036 &    0.0012 & 0.0030 \\

\end{tabular}

 	\caption{\em North Carolina 2010, calibration map}
     \end{subtable}
  \caption{\em Kuiper statistics \cite{tygert2021} for raking and BISG predictions on Florida
  (FL) and North Carolina (NC) registered voters using a calibration map and subsampling. 
  Each number corresponds to the total miscalibration over the 
  worst-case interval of predicted probabilities. Here, "worst-case" refers to the interval 
  that makes the absolute value of the total miscalibration as large as possible.
  Each entry of the table is also the difference between the maximum and 
  minimum cumulative miscalibration as plotted in Figure \ref{fig:calib_grid}.}  
   \label{t:kuiper}
\end{table}

In addition to reporting the above accuracy metrics, we measure calibration of raking and BISG 
predictions using the methods and software implementation of \cite{tygert2021}. 
Specifically, we graphically represent 
miscalibration with cumulative miscalibration plots (Figure \ref{fig:calib_grid}), and in 
Table \ref{t:kuiper} we report the corresponding Kuiper statistics. 
The Kuiper statistic is the total 
miscalibration over the worst-case interval of predicted probabilities, that is, the interval 
that makes the absolute value of the total miscalibration as large as possible. 
The Kuiper statistic is also the difference between the maximum 
and minimum of the cumulative miscalibration plotted in Figure \ref{fig:calib_grid}.


\section{Literature review}\label{a:lit_review}
Since its introduction in 2009, BISG has been widely used and has
inspired a large body of related work, including algorithm modifications and 
error analysis. Much of the work on BISG's shortcomings 
has focused on evaluating accuracy on labeled data sets. For example 
\cite{baines2014, zhang2018} used a data set of mortgage applications 
to demonstrate that BISG can overestimate racial disparities. 
Other studies, such as \cite{decter-frain2022b, decter-frain2022, imai2016}, 
have evaluated BISG's performance on data sets of registered voters. 
While these analyses provide a valuable contribution to understanding BISG,
their results combine errors from the conditional independence assumption
of BISG and differences in the distributions of the test set (registered voters,
or mortgage applicants) and the BISG factors (full U.S. population). 
In Appendix \ref{a:bisg_fl}, we isolated the impact of BISG's conditional 
independence assumption by fitting and testing BISG on the same labeled data set. 

To our knowledge, no prior research has been conducted on directly measuring the 
impact of BISG's independence assumption on its accuracy. Instead, 
researchers have relied on indirect evidence, such as using the accuracy of BISG 
on a labeled data set to demonstrate that the assumption does not result in 
excessively inaccurate predictions. Various potential sources of inaccuracy in the BISG assumption have been observed
in the literature. For example, \cite{imai2016, imai2022} identify issues including 
(i) within a given race/ethnicity, correlation between surname and wealth combined with
clustering of people of similar wealth may lead to correlation between geolocation and surname, 
(ii) relatives may live in close proximity to each other, and
(iii) ethnic groups within the same race/ethnicity (e.g., Indians and Chinese) have distinct
surnames and cluster geographically.

Other recent work has focused on how inaccuracy in BISG can impact 
downstream tasks such as regression analysis. For instance, among other 
contributions, \cite{chen2019} analyzes the mathematical sources of the overestimation 
of racial disparities that was observed empirically in \cite{baines2014, zhang2018}. 
Two authors of \cite{chen2019} build upon their analysis in \cite{kallus2022}
and observe that disparities are generally unidentifiable with only proxy 
information. They also propose methods for addressing such issues. 

BISG has not only inspired studies on its accuracy; much progress has 
been made on methodological improvements. These improvements include 
incorporating features in addition to surname and geolocation 
\cite{imai2016, imai2022, voicu2018, zhang2018, haas2019}, the use of 
Bayesian methods \cite{imai2022}, and addressing errors arising from the 
Census only publishing race/ethnicity distributions of common surnames 
\cite{imai2022}. 
There is also a vast literature on name-based 
prediction of race/ethnicity that does not necessarily use geographic 
information and has developed largely independently of BISG-related 
methods. See, for example, \cite{jain2022, sood2018, lee2017} for three 
recent approaches. A comprehensive review of this field can be found in 
\cite{jain2022}, but is beyond the scope of this paper. 

The use of known margins to calibrate BISG predictions has previously 
been used in healthcare applications. For example, in \cite{elliott2009}
the authors use multinomial logistic regression to transform standard BISG 
predictions to predictions for members of a national healthcare plan 
who self-report race/ethnicity. 
Their multinomial logistic regression uses labeled race information of members
of a national healthcare plan as the observation and BISG predictions as the 
predictors. Applying the fitted logistic transformation to BISG predictions then 
provides a healthcare-plan specific race/ethnicity prediction. 
Similar BISG adjustments are used in, for example, \cite{haas2019, branham2022}. 

This multinomial logistic correction is similar in spirit to the purpose of raking 
in this paper. In both cases, after adjusting BISG predictions, the race/ethnicity 
margin of new predictions matches exactly a known race/ethnicity distribution, 
in the case of \cite{elliott2009}, the margin of race/ethnicity of healthcare plan 
members who self-reported race/ethnicity. 
This strategy is not directly applicable to the problem we address in this paper
since that approach requires labeled race/ethnicity information that is usually
missing for registered voters. For example, in states such 
as New York, the surname by geolocation margin of registered voters is publicly 
available, but race/ethnicity information of individual registered voters is not available. 
It is possible to fit a multinomial logistic correction to our predictions by, say, 
starting with BISG predictions and solving a constrained optimization problem 
such that the margins of the predictions match the known margins. This can 
be solved with standard optimization packages. We did not implement this for this 
paper.\footnote{We implemented a similar correction that 
shifts predictions on the logistic scale, as opposed to raking, which multiplies 
probabilities. The results were similar to raking and given that raking
is a classical and standard tool, we focus on our raking-based method.}

In another healthcare-related BISG improvement, \cite{haas2019} builds a statistical 
model that incorporates demographic features including self-reported race/ethnicity 
in order to predict race/ethnicity of Medicare beneficiaries. This approach allows
flexible modeling that can relax BISG's independence assumption. 
As in \cite{elliott2009}, this approach also relies on labeled race information that is 
generally unavailable for the registered voter population. Only a handful of states 
provide any self-identified race/ethnicity, and when that data is provided, proxy 
methods are generally not required. 
It's possible that fitting a model in one state and applying it in another can improve
accuracy in race/ethnicity predictions. We leave that as an area of future research. 

\end{document}